\documentclass[10pt,conference]{IEEEtran}

\IEEEoverridecommandlockouts
\usepackage{graphicx}
\usepackage{subfigure}
\graphicspath{{./figure/}}
\usepackage{cite}
\usepackage{amsmath,amssymb,amsfonts}
\usepackage{algorithmic}
\usepackage{graphicx}
\usepackage{textcomp}
\usepackage{xcolor}
\usepackage{multirow}
\usepackage{booktabs}
\usepackage{framed}
\definecolor{gray}{gray}{0.9}
\usepackage{colortbl}
\usepackage[ruled,vlined,linesnumbered]{algorithm2e} 
\usepackage{url}
\usepackage{comment}
\usepackage{balance}

\newcommand{\redstar}{{\color{red}{$\bigstar$}}}
\newcommand{\cyanlozenge}{{\color{cyan}{$\blacklozenge$}}}
\newcommand{\yellowtriangle}{{\color{black!30!green}{$\blacktriangle$}}}

\begin{document}

\title{Measuring Discrimination to Boost Comparative Testing for Multiple Deep Learning Models
}

\author{
	Linghan Meng$^{1,2}$,  Yanhui Li$^{1,2,*}$, Lin Chen$^{1,2}$, Zhi Wang$^{1,2}$, Di Wu$^{3}$, Yuming Zhou$^{1,2}$, Baowen Xu$^{1,2}$ \\		
	1. State Key Laboratory for Novel Software Technology, Nanjing University, China \\
	 2. Department of Computer Science and Technology, Nanjing University, China\\
	3. Momenta, Suzhou, China\\
	\small 	
	\{menglinghan,wangz\}@smail.nju.edu.cn, \{yanhuili, lchen, zhouyuming, bwxu\}@nju.edu.cn, wudi@momenta.ai
	
	\thanks{* Yanhui Li is the corresponding and co-first author.}
}

\maketitle

\begin{abstract}

The boom of DL technology leads to massive DL models built and shared, which facilitates the acquisition and reuse of DL models.  
For a given task, we encounter multiple DL models available with the same functionality, which are considered as candidates to achieve this task. Testers are expected to \textit{compare} multiple DL models and select the more suitable ones w.r.t. the whole testing context. Due to the limitation of labeling effort, testers aim to select an efficient subset of samples to make an as precise rank estimation as possible for these models. 

To tackle this problem, we propose \textbf{S}ample \textbf{D}iscrimination based \textbf{S}election (\textbf{SDS}) to select efficient samples that could \textit{discriminate} multiple models, i.e., the prediction behaviors (right/wrong) of these samples would be helpful to indicate the trend of model performance.
To evaluate SDS, we conduct an extensive empirical study with three widely-used image datasets and 80 real world DL models. The experimental results show that, compared with state-of-the-art baseline methods, SDS is an effective and efficient sample selection method to rank multiple DL models.

\end{abstract}


\section{Introduction}
\label{sec:intro}

Deep learning (DL) supports a general-purpose learning procedure that discovers high-level representations of input samples with multiple layers of abstraction based on artificial neural networks (ANNs), which has shown significant advantages in establishing intricate structures of high dimensional data when tackling complex classification tasks \cite{lecun2015deep}. Along with increases in computation power \cite{You:2017} and data size \cite{geiger2012we}, DL technology achieves great success in constructing deeper layers of more effective abstraction to enhance classification performance, and has beaten human experts and traditional machine-learning technology in many areas \cite{pouyanfar2018survey}, including image recognition \cite{Krizhevsky2017ImageNet}, speech recognition \cite{hinton2012deep}, autonomous driving \cite{farabet2012scene}, playing Go \cite{silver2016mastering}, and so on. Meanwhile, concern about the reliability of DL models has been raised, which calls for novel testing techniques to deal with new DL testing scenarios and challenges. 

Most current DL testing techniques try to validate the quality of DL models in two testing scenarios:  
 debug testing and operational testing \cite{Frankl1998Evaluating,Li2019}. 
On the one hand, debug testing considers DL testing as a technology to improve reliability by finding faults\footnote{The faults of DL models are usually considered as the mismatching between the real labels and predicted labels of the input samples. } \cite{DeepRoad2018}, where various testing criteria (e.g., Neuron Activation Coverage \cite{DeepXplore2017} and Neuron Boundary Coverage \cite{DeepGauge2018}) have been proposed to generate or select error-inducing inputs which trigger faults. On the other hand, operational testing aims to make reliability assessment for DL models in the objective testing contexts. Li et al. proposed an effective operational testing technique to estimate the accuracy of a single DL model by constructing probabilistic models for the distribution of testing contexts \cite{Li2019}.

The boom of DL technology leads to DL models with ever-increasing functionality scale and complexity, i.e., complex DL models combine multi-function from multiple primitive DL models. Exposing code and data to build models and sharing model files (e.g., h5 files) boost the acquisition of DL models, which drive developers to build complex models by reusing available DL models achieving specific primitive functionality. One statistic in the previous study \cite{10.1145/3243734.3243757} indicates that more than 13.7\% of complex DL models on Github reuse at least one primitive DL model. On the positive side, this ``plug-and-play'' pattern \cite{sculley2015hidden} has greatly facilitated the construction and application of complex DL models. On the negative side, for a given DL task, it is tough to select suitable models because of the advent of numerous DL models constructed by mass developers.
These multiple models are produced by third-part developers and are trained on samples with different distributions. Therefore, their actual performance on the target application domain is not guaranteed, and they are needed to be tested.

These above points expedite the emergence of a new testing scenario \textit{``comparative testing''}, where testers may encounter multiple DL models with the same functionality built by different developers, all of which are considered as candidates to accomplish a specific task, and testers are expected to rank them to choose the more suitable models in the testing contexts. Generally speaking, comparative testing is different from current DL testing, i.e., debug and operational testing, in the following two points: 
\begin{itemize}
	\item The \textit{testing object} is multiple DL models instead of a single DL model;
	\item The \textit{testing aim} is comparing performances among multiple DL models instead of improving/assessing performances for a single DL model.   
\end{itemize}

\begin{figure}[t] 
\centering
\includegraphics[width=0.42\textwidth]{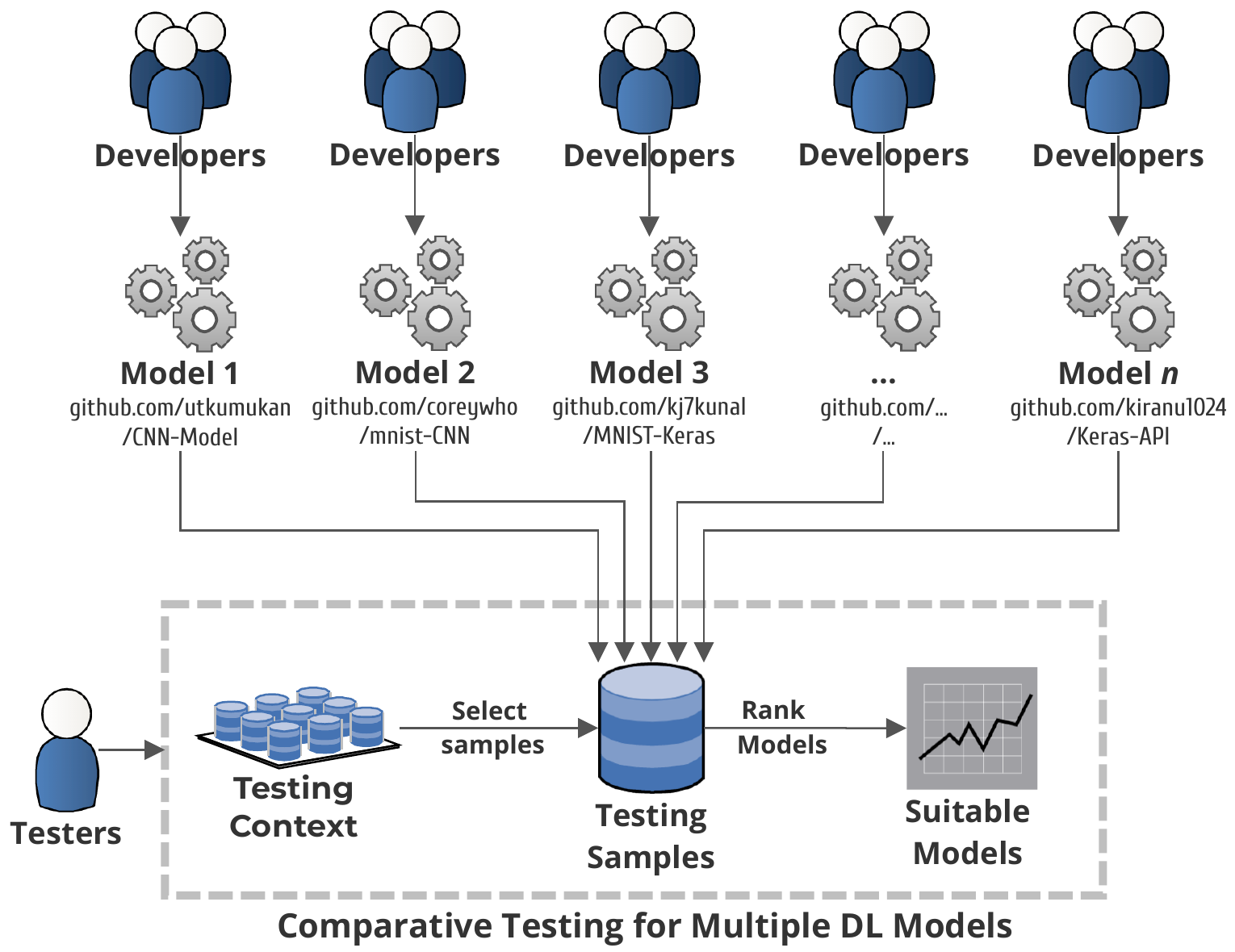}
\caption{An example of comparative testing scenarios with multiple real world DL models designed for written digit identification, all of which are trained on MNIST \cite{MNIST} dataset and available on GitHub. The testers need to evaluate and rank these DL models on the testing context. }
\label{moti}
\end{figure}

Figure~\ref{moti} shows an example of comparative testing scenarios considering multiple real world DL models available on GitHub. Hypothetically, in this scenario, the target application requires an implementation of written digit identification, and multiple candidate DL models are found to achieve this functionality. Testers are expected to compare the accuracy of written digit identification among these models and choose the more suitable ones to meet the requirements. As stated in many previous studies \cite{DeepGauge2018,feng2020,Li2019,9286133}, sample labeling is the bottle neck of testing resources for DL models, which spends much manpower and is time-consuming. Due to the limitation of labeling effort, testers can label only a very small part from the whole testing contexts. Therefore, as shown in Figure~\ref{moti},  
testers are asked to execute comparative testing by selecting and labeling a small but efficient subset of testing samples extracted from the testing context, and ranking multiple models based on their performance of the selected samples. 

As mentioned above, 
comparative testing brings out a new problem of DL testing: \textit{given limited labeling effort, how to select an efficient subset of samples (label and test them) to rank multiple DL models as precise as possible}? To tackle this problem, we propose a novel algorithm named \textbf{S}ample \textbf{D}iscrimination based \textbf{S}election (\textbf{SDS}) to measure the sample discrimination and select samples with higher discrimination. 
The main idea of our algorithm is to focus on efficient samples that could \textit{discriminate} multiple models, i.e., the prediction behaviors (right/wrong) of these samples would be helpful to indicate the trend of model performance. Specifically, SDS combines two aspects of technical thoughts: majority voting \cite{sagi2018ensemble} in ensemble learning and item discrimination \cite{ebel1954procedures} in test analysis, which are introduced to estimate the sample discrimination with the lack of actual labels (details are in Section~\ref{sec:method}).

We evaluate our approach on three widely-used image datasets MNIST \cite{MNIST}, Fashion-MNIST \cite{xiao2017fashion}, and CIFAR-10 \cite{cifar_10}, each of which contains 10000 testing samples. To simulate the comparative testing scenarios where multiple DL modes are developed/submitted for the same task (e.g., digital identification with MNIST and clothing classification with Fashion-MNIST), we introduce totally 80 models from GitHub, including 28 models for MNIST, 25 for Fashion-MNIST, and 27 for CIFAR-10.
To assess the performance of SDS, we introduce three sample selection methods as the baselines: one state-of-the-art method from debug testing (DeepGini at ISSTA'2020 \cite{feng2020}), one state-of-the-art method from operational testing (CES at FSE'2019 \cite{Li2019}) and the simple random selection (SRS). The experimental results indicate that our algorithm SDS is an effective and efficient sample selection method for comparative testing to solve the problem ``\textit{ranking multiple DL models under limited labeling efforts}''.

Our study makes the following contributions:

\begin{itemize}
	\item \textbf{Dimension.} This study opens a new dimension of DL testing ``comparative testing'' for DL models, which focuses on comparing multiple DL models instead of improving/assessing a single DL model.  

	\item \textbf{Strategy.} This paper proposes a novel selection method SDS to measure the discrimination of samples and select samples with higher discrimination to rank multiple DL models. 
	\item \textbf{Study.} This paper contains an extensive empirical study of 80 models with three datasets containing 10000 testing inputs. 
	The experimental results indicate that compared with the baseline methods, SDS is an effective and efficient sample selection method for comparative testing.  
\end{itemize}

The rest of this paper is organized as follows. 
In Section~\ref{sec:mainidea}, we introduce a motivation example to show the difference between comparative testing and debug/operational testing. In Section~\ref{sec:method}, we present a detailed description of our algorithm SDS. In Section~\ref{sec:expset}, we present our experimental settings, including studied datasets and models, baseline methods, research questions, and so on. Section~\ref{sec:expres} explains experimental results and discoveries. Section~\ref{sec:dis} further discusses some important experimental details. Sections \ref{sec:thread} and \ref{sec:rel} are threats to validity and related works, respectively. Section~\ref{sec:con} presents the conclusion of our paper.

\section{The Motivation Example}

\label{sec:mainidea}

As we mentioned in Section \ref{sec:intro}, the aim of comparative testing is comparing the performances of multiple models. Here we introduce an example to show the differences between comparative testing and debug/operational testing.

Figure~\ref{moti2} presents an example of comparative testing scenarios containing six testing samples $s_1,\dots,s_6$ and three DL models $\mathcal{M}_1, \mathcal{M}_2, \mathcal{M}_3$, with the prediction results of samples predicted by models. 
\checkmark/$\times$ indicates that the prediction results of these models running against samples are right/wrong (i.e., the predicted labels are identical/different with the actual ones). By calculating the numbers of \checkmark/$\times$, we can obtain the accuracies of three models, i.e., $\frac{4}{6}$ for $\mathcal{M}_1$, $\frac{3}{6}$ for $\mathcal{M}_2$, and  $\frac{2}{6}$ for $\mathcal{M}_3$, respectively. As a result, the actual rank of accuracies ($Acc(\,)$) for these models is 
$Acc(\mathcal{M}_1)>Acc(\mathcal{M}_2)>Acc(\mathcal{M}_3)$. As shown in Figure~\ref{moti2}, we have the following observations: 

\begin{itemize}
	\item As only six samples are considered, we can easily find that the most efficient subset to indicate the actual rank of these models is $S^*=\{s_1,s_2\}$: $\mathcal{M}_1$ has two \checkmark under $S^*$, $\mathcal{M}_2$ has one, and $\mathcal{M}_3$ has none. We can obtain the same rank of models for the accuracies ($Acc'(\,)$) w.r.t. $S^*$: 		
	$Acc'(\mathcal{M}_1)=\frac{2}{2}>Acc'(\mathcal{M}_2)=\frac{1}{2}>Acc'(\mathcal{M}_3)=\frac{0}{2}$.
	\item $S^*$ is not the target sampling subset in operational testing, as it assesses model performance imprecisely: the accuracy $Acc'(\mathcal{M}_1)/Acc'\mathcal{M}_2)/Acc'(\mathcal{M}_3)$ under $S^*$ is $\frac{2}{2}$/$\frac{1}{2}$/$\frac{0}{2}$, which is much different from the actual $\frac{4}{6}$/$\frac{3}{6}$/$\frac{2}{6}$. 
	\item $S^*$ is also not the target sampling subset in debug testing. Debug testing would consider $s_5$ with the highest priority, since it triggers the mismatching behaviors of all models. 
\end{itemize}

These observations indicate that the differences of aims between comparative testing and debug/operational testing lead to the different sampling priority. In comparative testing, we focus on the samples that could \textit{discriminate} multiple models, e.g., $s_1$ and $s_2$ in Figure 2.  
In the next section, we will introduce a novel algorithm to measure sample discrimination and select samples with higher discrimination.

\begin{figure}[t]
\centering
\begin{tabular}{|c|c|c|c|c|c|c|c|}
\hline
\multirow{2}*{No.} & \multicolumn{6}{c|}{Prediction results} & \multirow{2}*{Acc}\\
\cline{2-7} 
 & $s_1$ & $s_2$ & $s_3$ & $s_4$ & $s_5$ & $s_6$  & \\
\hline 
\hline
$\mathcal{M}_1$ & \checkmark  & \checkmark  & \checkmark  & \checkmark & $\times$ & $\times$ & 4/6 \\
\hline 
$\mathcal{M}_2$ & \checkmark & $\times$  & $\times$ &  \checkmark& $\times$ & \checkmark  & 3/6  \\
\hline 
$\mathcal{M}_3$ & $\times$  & $\times$ &  \checkmark & $\times$ & $\times$  & \checkmark & 2/6 \\
\hline 
\end{tabular}
\caption{An example of six testing samples with prediction results under three DL models. \checkmark and $\times$ show the prediction result: right and wrong. }
\label{moti2}
\end{figure}

\section{Methodology} 
\label{sec:method}

In this section, we present the detailed description of our approach. First, we present the studied problem. Next, we show an algorithm named Sample Discrimination based Selection (SDS) to measure the sample discrimination and select samples with higher discrimination.

\subsection{The Studied Problem}
\label{problem}

We first introduce some symbols and definitions, which are helpful for readers to understand the rest of our paper. 

\textbf{Definition 1 (DL models).} A DL model $\mathcal{M}$ is usually regarded as an implementation of complex classification task based on the layer structure of artificial neural networks, which achieves a function mapping the high dimensional samples $s$ (e.g., a gray value matrix for figures) to labels $\mathcal{L}$ in a given label set $S_\mathcal{L}=\{\mathcal{L}_1,\mathcal{L}_2,...,\mathcal{L}_c\}$: $\mathcal{M}(s)\in  \mathcal{S}_\mathcal{L}$.

\textbf{Definition 2 (Accuracy).} A DL model $\mathcal{M}$ is tested under the testing context $\mathcal{C}_t$ containing samples $s$. Let $\mathcal{M}(s)$ and $\mathcal{L}(s)$ be the predicted label generated by $\mathcal{M}$ and the actual label of $s$, respectively. The accuracy $Acc(\mathcal{M},\mathcal{C}_{t})$ of $\mathcal{M}$ w.r.t. $\mathcal{C}_{t}$ is defined as follows: 
\[Acc(\mathcal{M},\mathcal{C}_{t})=\frac{|\{s|s\in \mathcal{C}_{t}, \mathcal{M}(s)=\mathcal{L}(s) \}|}{|\mathcal{C}_{t}|}\]

We introduce accuracy as the main indicator to measure the performance for comparing multiple DL models, as it has been widely used in evaluating the performance of DL models \cite{Li2019,DeepMutation}.  Based on above definitions and symbols, we present the studied problem ``given limited labeling effort, for multiple DL models, tester aim to select an efficient subset of samples (label and test them) to rank these models as precise as possible'' specifically:

\begin{framed}
\noindent Problem. $\mathcal{M}_1,\mathcal{M}_2,\cdots,\mathcal{M}_n$ are tested under the testing context $\mathcal{C}_{t}$. and all samples $s$ in $\mathcal{C}_{t}$ are unlabeled. Given limited labeling effort $\mathcal{E}$ ($\mathcal{E}\ll|\mathcal{C}_{t}|$), the task is to select and label an efficient subset $\mathcal{C}_{r}$ ($|\mathcal{C}_{r}|=\mathcal{E}$)  
from $\mathcal{C}_{t}$,
and employ the results (i.e., $Acc(\mathcal{M}_i,\mathcal{C}_{r}))$ on $\mathcal{C}_{r}$ to estimate the \textbf{rank} of model performance (i.e., $Acc(\mathcal{M}_i,\mathcal{C}_{t})$)
on the whole testing context $\mathcal{C}_{t}$, with an as small rank error as possible. 
	
\end{framed}

\subsection{Sample Discrimination based Selection}
           \label{sec:alog}
 
 As shown in the motivation example, comparative testing need samples that could discriminate the multiple models. In this subsection, we propose a novel algorithm named \textbf{S}ample \textbf{D}iscrimination based \textbf{S}election (SDS) to measure the sample discrimination and select samples with higher discrimination. Generally, SDS combines two aspects of technical thoughts:

\begin{algorithm}[t] 
\footnotesize 
            \caption{\textbf{S}ample
            \textbf{D}iscrimination based \textbf{S}election $\mathrm{SDS}(S_\mathcal{M},\mathcal{C}_t,S_\mathcal{L})$}
            \label{alg:Framwork} 
            \KwIn{the set of DL models $S_\mathcal{M}=\{\mathcal{M}_1,\mathcal{M}_2,\cdots,\mathcal{M}_n\}$, the testing context $\mathcal{C}_t=\{s_1,s_2,\cdots,s_m\}$ with unlabeled sample $s_i$, and the label set $S_\mathcal{L}=\{\mathcal{L}_1,\mathcal{L}_2,...,\mathcal{L}_c\}$.}
            \KwOut{the subset $\mathcal{C}_r$ with $\mathcal{C}_r\subset \mathcal{C}_t$ and $|\mathcal{C}_r|=\mathcal{E}$.}
            initialize $\mathcal{C}_r=\emptyset$\;
            initialize an array $\mathcal{A}_d[1\dots m]$: $\mathcal{A}_d[i]=0$, $1\leq i\leq m$\;
            initialize a two dimensional ($n\times m$) array $\mathcal{A}_p$ that stores the prediction matrix of $n$ models on $m$ samples, i.e., $\mathcal{A}_p[i][j]$ ($1\leq i\leq n$, $1\leq j\leq m$ ) indictors that $\mathcal{M}_i$ predicts $s_j$ as the label $\mathcal{A}_p[i][j]$, with $\mathcal{A}_p[i][j]=\text{null}$\; 
            initialize an array $\mathcal{A}_f[1\dots c]$ that stores the frequency of labels in the prediction results with $\mathcal{A}_f[k]=0$, $1\leq k \leq c$\;  
            initialize an array $\mathcal{A}_v[1\dots m]$ that stores the voting labels of $m$ samples with $\mathcal{A}_v[j]=\text{null}$, $1\leq j \leq m$\; 
            initialize an array $\mathcal{A}_s[1\dots n]$ that stores the scores of $n$ models with $\mathcal{A}_s[m]=0$, $1\leq i \leq n$\;          
            \For(\tcp*[f]{\color{red}{1: Extract prediction results}}){$i=1$ \KwTo $n$}{
            	\For{$j=1$ \KwTo $m$}{
            		run $\mathcal{M}_i$ on $s_j$ and get the prediction label $\mathcal{L}_{p}\in S_\mathcal{L}$\;
            		assign the prediction label to $\mathcal{A}_p[i][j]$: $\mathcal{A}_p[i][j]=\mathcal{L}_{p}$\;
            	}
            }
            \For(\tcp*[f]{\color{red}{2: Vote for sample labels}}){$j=1$ \KwTo $m$}{ 
            	\For{$k=1$ \KwTo $c$}{
            		Count the frequency of $\mathcal{L}_k$ in the $n$ prediction results $\mathcal{A}_p[:,j]$ of sample $s_j$: $\mathcal{A}_f[k]=\mathsf{freq}(\mathcal{L}_k,\mathcal{A}_p[:,j])$ \;
            	}
            	Use the majority voting results as the actual labels:   $k^* =\mathop{\arg\max}_{1\leq k\leq c}\{\mathcal{A}_f[k]\}$,  $\mathcal{A}_v[j]=\mathcal{L}_{k^*}$\;
            	
            }     
             \For(\tcp*[f]{\color{red}{3: Classify top/bottom models}}){$i=1$ \KwTo $n$}{
             	initialize $score=0$\;
            	\For{$j=1$ \KwTo $m$}{
            		\If{$\mathcal{A}_p[i][j]=\mathcal{A}_v[j]$}{
            			$score = score +1$\; 
            		} 
            	}
            	$\mathcal{A}_s[i]=score$\; 
            }	   
           Sort $n$ DL models in descending order by $\mathcal{A}_s[i]$\;
           Select the top and the bottom 27\% models into $S_t$ and $S_b$, respectively\; 
           \For(\tcp*[f]{\color{red}{4: Compute sample discrimination}}){$j=1$ \KwTo $m$}{
           		initialize $discrimination=0$\; 
           		\For{$i=1$ \KwTo $n$}{
           			\If{$\mathcal{M}_i\in S_t$}{
           				\If{$\mathcal{A}_p[i][j]=\mathcal{A}_v[j]$}{
            			$discrimination = discrimination +1$\; 
            		} 
           			}
           			\ElseIf{$\mathcal{M}_i\in S_b$}{
           				\If{$\mathcal{A}_p[i][j]=\mathcal{A}_v[j]$}{
            			$discrimination = discrimination +(-1)$\; 
            		} 
           			}
           		}
           		$\mathcal{A}_d[j]=discrimination/|S_t|$
           }
            Sort $m$ samples by their discrimination $\mathcal{A}_d[j]$ in descending order\tcp*[r]{\color{red}{5: select with higher discrimination}} 
            Select the top 25\% samples into the candidate set $S_c$\;
            Randomly select $\mathcal{E}$ samples from $S_c$ into $\mathcal{C}_r$\;  
            \Return $\mathcal{C}_r$\;
    \end{algorithm}

\begin{itemize}
	\item Majority voting \cite{sagi2018ensemble}. Majority voting is a simple weighting method in ensemble learning, which selects the class with the most votes as the final decision. As our algorithm has the precondition that all samples are unlabeled, we employ majority voting as a procedure to deal with the lack of actual labels, i.e., for a given sample, we choose the predicted label with the most models as the estimation of the actual label. 
	\item Item discrimination \cite{ebel1954procedures}. Item discrimination is an indicator to describe to what extent test items can discriminate between good and poor students, which is widely used in test analysis\footnote{https://www.medsci.ox.ac.uk/divisional-services/support-services-1/learning-technologies/faqs/what-do-difficulty-correlation-discrimination-etc-in-the-question-analysis-mean}. We introduce the idea of item discrimination to measure sample discrimination, i.e., estimate discrimination by calculating the difference performance between good and bad models under each sample. 	 	
\end{itemize}

Specifically, given multiple DL models  $\mathcal{M}_1,\mathcal{M}_2,\cdots,\mathcal{M}_n$, the testing context $\mathcal{C}_t=\{s_1,s_2,\cdots,s_m\}$ with unlabeled samples $s_i$, the label set $S_\mathcal{L}=\{\mathcal{L}_1,\mathcal{L}_2,...,\mathcal{L}_c\}$, and the labeling effort $\mathcal{E}$, SDS is composed of the following five steps, as shown in Algorithm 1.

Step 1: Extract prediction results. We run multiple DL models against the testing context (line 9).  For model $\mathcal{M}_i$  and sample $s_j$, we record the predicted label $\mathcal{L}_p$ in the element $\mathcal{A}_p[i][j]$ of the prediction matrix $\mathcal{A}_p$ (line 10). 

Step 2: Vote for estimated labels. For any sample $s_j$, we compute the frequency of predicted labels created by multiple models (line 13). We choose the predicted label with the max frequency, i.e., majority voting, as the estimated label (line 14), which is the basic of the following steps.   

Step 3: Classify top/bottom models. We employ the voted labels to score the predicted results of DL models on samples one by one, if the predicted label equals to the voted label, we add one score for the current model (line 19). After we go through all the samples, we obtain an estimated score for this model. We sort $n$ models in descending order by their estimated score (line 21).  
	According to the classification in \cite{ebel1954procedures}, we classify $n$ models into three classes (line 22): \textit{top} class containing the top 27\% models, \textit{bottom} class containing the bottom 27\% models, and \textit{other} class containing other models. 
	
Step 4: Compute sample discrimination.  We employ difference performance of models in top/bottom class to calculate discrimination. Specifically, for each sample $s_j$, the value of discrimination is the number of models with right prediction in top class minus the number in bottom class (line 23-31). Intuitively, if the number in top class is much larger than the number in bottom class, the result of this sample is more identical with the rank, i.e., it would be helpful to estimate the rank of model performance. 
Finally, we normalize and
store the sample discrimination (line 32).  

Step 5: We consider the samples with higher discrimination as the ones which are more helpful to rank multiple DL models. To eliminate the effects of outlier samples with higher discrimination, we introduce random selecting instead of direct selecting from higher discrimination to lower discrimination. Specifically, we choose 25\% as the cutoff point to construct the subset of samples with higher discrimination since quartering is common for dataset partition in software engineering \cite{Surprise_Adequacy}, i.e., we consider the top 25\% samples as the candidates (line 34) and randomly select samples from them according to the given labeling effort (line 35). 
	
Figure~\ref{examofalg} shows an example of SDS running on four DL models $\mathcal{M}_1$, \dots, $\mathcal{M}_4$ with the testing context containing four samples $s_1,s_2,\cdots,s_4$, which are classified into three classes \redstar, \yellowtriangle, and \cyanlozenge. Four subfigures show the running results of the first four steps\footnote{As step 5 is easy to understand, we omit its running here.} of SDS, respectively, where the entries with a gray background indicates the target information obtained in each step. Next, we describe the subfigures one by one. 

\begin{itemize}
	\item Figure~\ref{suba} shows that SDS constructs the $4\times 4 $ prediction matrix, where \redstar, \yellowtriangle, and \cyanlozenge\, are the predicted labels.
	\item Figure~\ref{subb} presents that SDS employs majority voting to obtain the estimation of actual labels. For example, for $s_1$, three models predict it as \redstar, and one as \cyanlozenge. Therefore, SDS adds \redstar\, as its estimated label.
	\item Figure~\ref{subc} presents that SDS estimates the scores of models based on estimated labels, e.g., since $\mathcal{M}_3$ have three right and one wrong prediction, $\mathcal{M}_3$ is scored 3; and SDS classify $M_1$ into the top class and $M_2$ into the bottom class.  	     
	\item  Figure~\ref{subd} shows SDS counts the number of models with right prediction in top class minus the number in bottom class, e.g., for $s_2$, both  $\mathcal{M}_1$ and $\mathcal{M}_2$ predict right, the discrimination of $s_2$ is $1+(-1)=0$.
\end{itemize}

\begin{figure}[t]
	\centering
	\subfigcapskip=4pt
	\subfigure[Extract prediction results]{
	\label{suba}
	\scalebox{0.8}{
\begin{tabular}{|c|c|c|c|c|p{0.4cm}|}
\hline
 & \multicolumn{4}{c|}{Prediction} & \multirow{2}*{S}\\
\cline{2-5} 
No. & $s_1$ & $s_2$ & $s_3$ & $s_4$ & \\
\hline 
\hline
$\mathcal{M}_1$ & \cellcolor{gray}\redstar & \cellcolor{gray}\yellowtriangle & \cellcolor{gray}\cyanlozenge & \cellcolor{gray}\redstar & ?\\
\hline 
$\mathcal{M}_2$ & \cellcolor{gray}\cyanlozenge & \cellcolor{gray}\cellcolor{gray}\yellowtriangle & \cellcolor{gray}\redstar & \cellcolor{gray}\redstar & ?\\
\hline 
$\mathcal{M}_3$ & \cellcolor{gray}\redstar & \cellcolor{gray}\yellowtriangle & \cellcolor{gray}\cyanlozenge & \cellcolor{gray}\yellowtriangle & ?\\
\hline 
$\mathcal{M}_4$ & \cellcolor{gray}\redstar & \cellcolor{gray}\redstar & \cellcolor{gray}\cyanlozenge & \cellcolor{gray}\redstar & ?\\
\hline 
\hline 
L & ? & ? & ? & ? & \\
\hline 
D & ? & ?& ? & ? & \\
\hline 
\end{tabular}}
}
\subfigure[Vote for sample labels]{
\label{subb}
\scalebox{0.8}{\begin{tabular}{|c|c|c|c|c|p{0.4cm}|}
\hline
 & \multicolumn{4}{c|}{Prediction} & \multirow{2}*{S}\\
\cline{2-5} 
No. & $s_1$ & $s_2$ & $s_3$ & $s_4$ & \\
\hline 
\hline
$\mathcal{M}_1$ & \redstar & \yellowtriangle & \cyanlozenge & \redstar & ?\\
\hline 
$\mathcal{M}_2$ & \cyanlozenge & \yellowtriangle & \redstar & \redstar & ?\\
\hline 
$\mathcal{M}_3$ & \redstar & \yellowtriangle & \cyanlozenge & \yellowtriangle & ?\\
\hline 
$\mathcal{M}_4$ & \redstar & \redstar & \cyanlozenge & \redstar & ?\\
\hline 
\hline 
L & \cellcolor{gray}\redstar & \cellcolor{gray}\yellowtriangle & \cellcolor{gray}\cyanlozenge & \cellcolor{gray}\redstar & \\
\hline
D & ? & ?& ? & ? & \\
\hline 
\end{tabular}}
}\\
\subfigure[Classify top/bottom models]{
\label{subc}
\scalebox{0.8}{
\begin{tabular}{|c|c|c|c|c|p{0.4cm}|}
\hline
 & \multicolumn{4}{c|}{Prediction} & \multirow{2}*{S}\\
\cline{2-5} 
No. & $s_1$ & $s_2$ & $s_3$ & $s_4$ & \\
\hline 
\hline
$\mathcal{M}_1$ & \checkmark  & \checkmark  & \checkmark  & \checkmark & \cellcolor{gray}4:T \\
\hline 
$\mathcal{M}_2$ & $\times$ & \checkmark  & $\times$ & \checkmark & \cellcolor{gray}2:B \\
\hline 
$\mathcal{M}_3$ & \checkmark  & \checkmark  & \checkmark  & $\times$ & \cellcolor{gray}3 \\
\hline 
$\mathcal{M}_4$ & \checkmark & $\times$ & \checkmark & \checkmark & \cellcolor{gray}3 \\
\hline 
\hline 
L & \redstar & \yellowtriangle & \cyanlozenge & \redstar & \\
\hline
D & ? & ?& ? & ? & \\
\hline 
\end{tabular}}
}
\subfigure[Compute sample discrimination]{
\label{subd}
\scalebox{0.8}{
\begin{tabular}{|c|c|c|c|c|p{0.4cm}|}
\hline
 & \multicolumn{4}{c|}{Prediction} & \multirow{2}*{S}\\
\cline{2-5} 
No. & $s_1$ & $s_2$ & $s_3$ & $s_4$ & \\
\hline 
\hline
$\mathcal{M}_1$ & 1  & 1 & 1  & 1 & T \\
\hline 
$\mathcal{M}_2$ & - & -1  & - & -1 & B \\
\hline 
$\mathcal{M}_3$ & -  & -  & -  & - & - \\
\hline 
$\mathcal{M}_4$ & - & - & - & - & - \\
\hline 
\hline 
L & \redstar & \yellowtriangle & \cyanlozenge & \redstar & \\
\hline
D & \cellcolor{gray}1 & \cellcolor{gray}0 & \cellcolor{gray}1 & \cellcolor{gray}0 & \\
\hline
\end{tabular}}
}
\caption{An example of SDS running on four DL models $\mathcal{M}_1$, \dots, $\mathcal{M}_4$ with the testing context containing four samples ($s_1,s_2,\cdots,s_4$), which are classified into three classes (\redstar, \yellowtriangle, and \cyanlozenge).  Due to the limited space, some abbreviations are used in the subfigures (L: estimated label, S: score, D: discrimination, T: top 27\% class, B: bottom 27\% class).}
\label{examofalg}
\end{figure}

\section{EXPERIMENTAL SETUPS}
\label{sec:expset}

In this section, we present the experimental setup to evaluate the performance of SDS. 

\begin{table*}[t]
	\caption{The detailed description of the studied 80 DL models, including \textbf{28} for MNIST, \textbf{25} for Fashion-MNIST, and \textbf{27} for CIFAR-10.}
	\label{modeldetail} 
	\centering 
\scalebox{0.95}{	
	\footnotesize 
	\begin{tabular}{|l|l|p{8.35cm}|ll|l|}
		\hline
		\multirow{2}*{Dataset} &	Model            & \multirow{2}*{GitHub Website}                                                                  & \multicolumn{2}{l|}{Model Structure}  &
		\multirow{2}*{Actual Accuracy}                     \\ 
		\cline{4-5}
		&No. & & Layers & Params  &         \\
		\hline
		\hline
		\multirow{10}*{MNIST} 	& 1,2           & https://github.com/Rowing0914/simple\_CNN\_mnist                                       & 8      & 1199882  & 0.9889-0.9905       \\
		& 3-9           & https://github.com/utkumukan/CNN-Model                                                 & 14     & 206826 & 0.9858-0.9916         \\
		& 10            & https://github.com/11510880/Keras\_model\_MNIST\_99.66-                                       & 31     & 696402    & 0.9912      \\
		& 11            & https://github.com/nanguoyu/MNIST\_Keras\_CNN                                                 & 12     & 600810     & 0.992      \\
		& 12            & https://github.com/coreywho/mnist-CNN                                                        & 9      & 79280     & 0.9919       \\
		& 13            & https://github.com/kj7kunal/MNIST-Keras                                                      & 20     & 330730     & 0.9925      \\
		& 14            & https://github.com/gee842/MNIST-Models                                                       & 19     & 327242     & 0.9959      \\
		& 15            & https://github.com/Aishuvenkat09/Predictions-using-Mnist-Model                               & 8      & 1199882    & 0.9915      \\
		& 16-22         & https://github.com/keras-team/keras                                                & 8      & 151306-1199882 & 0.9883-0.9922 \\
		& 23-28         & https://github.com/avicorp/AmountRecognition/                                          & 8-9    & 7218-444986  & 0.8818-0.9853   \\
		\hline
		\hline
		\multirow{5}*{Fashion-MNIST} & 1,2   & https://github.com/avicorp/AmountRecognition/                                          & 31     & 258826   & 0.9096-0.9254       \\
		& 3     & https://github.com/fwsdonald/classification-of-Fashion-Mnist                                  & 12     & 329962   & 0.9315       \\
		& 4,5   & https://github.com/Sukhman75/Tensorflow\_Keras\_fashion\_mnist                         & 11-13  & 356234-1199882 & 0.9195-0.9335 \\
		& 6-20  & https://github.com/zsoltzombori/keras\_fashion\_mnist\_tutorial                        & 7-31   & 693962-258826  & 0.9046-0.9338 \\
		& 21-25 & https://github.com/zk31601102/FGSM-fashion-mnist                                       & 7      & 931080-1256080 & 0.9009-0.9273 \\
		\hline
		\hline
		\multirow{11}*{CIFAR-10} & 1-3           & https://github.com/kiranu1024/Keras-API                                                   & 18-72  & 274442-1250858  & 0.7238-0.8000\\
		& 4,5           & https://github.com/uchidama/CIFAR10-Prediction-In-Keras                                & 18-72  & 274442-1250858 & 0.7747-0.8527 \\
		& 6-9           & https://github.com/Ken-Leo/BIGBALLONcifar-10-cnn                                       & 8-65   & 62006-39002738 & 0.7149-0.7529 \\
		& 10,11         & https://github.com/hemrajchauhan/CIFAR10\_Keras                                        & 12     & 1250858  & 0.7701-0.7926      \\
		& 12-15         & https://github.com/night18/cifar-10-AlexNet                                            & 11-13  & 1248554-2883178 & 0.7009-0.7336 \\
		& 16            & https://github.com/sahilunagar/CIFAR-10-image-classification-using-CNN-model-in-keras         & 15     & 2122186    & 0.7362     \\
		& 17            & https://github.com/saranshmanu/CIFAR-Image-Classification                                     & 19     & 781992  & 0.7536        \\
		& 18,19         & https://github.com/GodfatherPacino/CNN\_CIFAR                                         & 8-19   & 2915114-4210090 & 0.7091-0.7944 \\
		& 20            & https://github.com/sonamtripathi/simple\_cnn\_model\_keras\_cifar10\_dataset                  & 12     & 1250858   & 0.7859      \\
		& 21            & https://github.com/percent4/resnet\_4\_cifar10                                                & 72     & 274442   & 0.7622       \\
		& 22-27         & https://github.com/BIGBALLON/cifar-10-cnn                                              & 113    & 470218 & 0.7748-0.8081 \\      
		\hline  
	\end{tabular}
	}
\end{table*}

\subsection{Studied Dataset and Models}

We introduce three widely used datasets MNIST \cite{MNIST}, Fashion-MNIST \cite{xiao2017fashion}, and  CIFAR-10 \cite{cifar_10} to conduct our experiments.  
MNIST is a dataset of handwritten digit images with 60000 training samples and 10000 testing samples. Samples in MNIST are $28\times28$ pixel grayscale images to denote handwritten digits from 0 to 9.
	Fashion-MNIST is similar with MNIST, containing 60000 training samples and 10000 testing samples which are $28\times28$ pixel grayscale images to describe ten types of clothing.
CIFAR-10 contains 60000 $32\times32$ pixel color images (50000 for training and 10000 for testing), which are equally distributed into 10 classes, e.g., cat, dog, ship, and truck. In summary, each dataset supports 10000 testing samples, which are considered as the testing context in the following experiment. 

For these three datasets, we extract a large amount of (80) models on Github, 28 models for MNIST, 25 for Fashion-MNIST, and 27 for CIFAR-10, respectively. To simulate the different implements of the same tasks, we choose these DL models with different stars (from a few to tens of thousands) on Github, different model structures, and different accuracies.   
For each model, if the model files (e.g., saved as h5 file) are provided in the repository on GitHub, we reuse them directly; otherwise, we employ the code and data provided to train the studied models. Table~\ref{modeldetail} presents the detailed description of these studied models with their GitHub repositories, parameters of model structure, and actual accuracies in the testing context. As shown in Table~\ref{modeldetail}, some of studied models come from the same repository, we put them together and provide the minimum and maximum values of layers, parameters, and accuracies among them.

\subsection{Experimental Settings}

This section will describe some details in the experimental settings in the following aspects.

\textbf{Sampling Size.} As mentioned above, the labeling effort is the bottle neck, i.e., tester are limited to label only the very small percent of testing samples. Following the experiment design in \cite{Li2019},  we focus on results on each sample size from 35 to 180 with intervals of 5 (i.e., 35, 40, \dots, 180), which are 0.35\%-1.8\% samples selected from the whole testing context.

\textbf{Baseline Method.}
Given multiple DL models, our goal is to rank the performance of these models by selecting and labeling a discriminative subset. It's worth pointing out that, as comparative testing is a new testing scenario proposed in this paper, there are not existing baseline methods. To clarify the performance of our method, we conduct comparative experiments with three baselines: two state-of-the-art sample selection methods (CES at FSE'2019 \cite{Li2019} and DeepGini at ISSTA'2020 \cite{feng2020}) in current DL testing and random selection. 

\begin{itemize}
	\item CES: Li et al. proposed an effective method named CES to select samples for DL testing to assess the accuracy of the single DL model. We choose CES as a baseline as it also aims to select representative subsets of sample and reduce the labeling costs. Since CES runs based on the single model, given $n$ DL models, CES may construct $n$ selections of samples for $n$ models, respectively. Here, we introduce the best of $n$ selections (i.e., choose the subset that gets the highest performance of ranking) as the result generated by CES, which is a stronger baseline to show the advance of our method.  		 

	\item DeepGini: Feng et al. proposed a technique called DeepGini to help prioritize testing DL models, which measures the likelihood $\xi$ of misclassification by calculating the set impurity of prediction probabilities for multiple classification. 
	DeepGini supports a deterministic baseline method, i.e., it sorts the test samples according to the calculated likelihood $\xi$ and selects the samples according to sampling size. As SDS and CES are  with randomness, we combine random selection and DeepGini to construct a new baseline, in which we perform random sampling in the first 25\% (the same cutoffs in SDS) samples  according to the rank of $\xi$. To differentiate these two baseline methods, we call the the former deterministic DeepGini (DDG), and the latter random DeepGini (RDG).

	\item SRS: Simple Random Sampling (SRS) is a basic method for subset selection, which is used as baseline for many studies \cite{Li2019}. We randomly select a subset from the testing set and test the ranking performance of this subset.
\end{itemize}

We implement SDS and baseline methods in python 3.6.3 with the frameworks including Tensorflow 2.3.0 and Keras 2.4.3. Our experiments are performed on a Ubuntu 18.04 server with 8 GPU cores ``Tesla V100 SXM2 32GB''. 
We provide the replication package including the detailed description of our proposed methods SDS and source code online (see Section \ref{rep}). 

\textbf{Repetition.}
As SDS and several baseline methods are with randomness, we conduct the experiment 50 times and report the average of calculated results.

\begin{figure*}[t]
	\centering
	\subfigure[Spearman coefficient of MNIST]{
		\label{Main-M-S} 
		\includegraphics[width=0.31\textwidth]{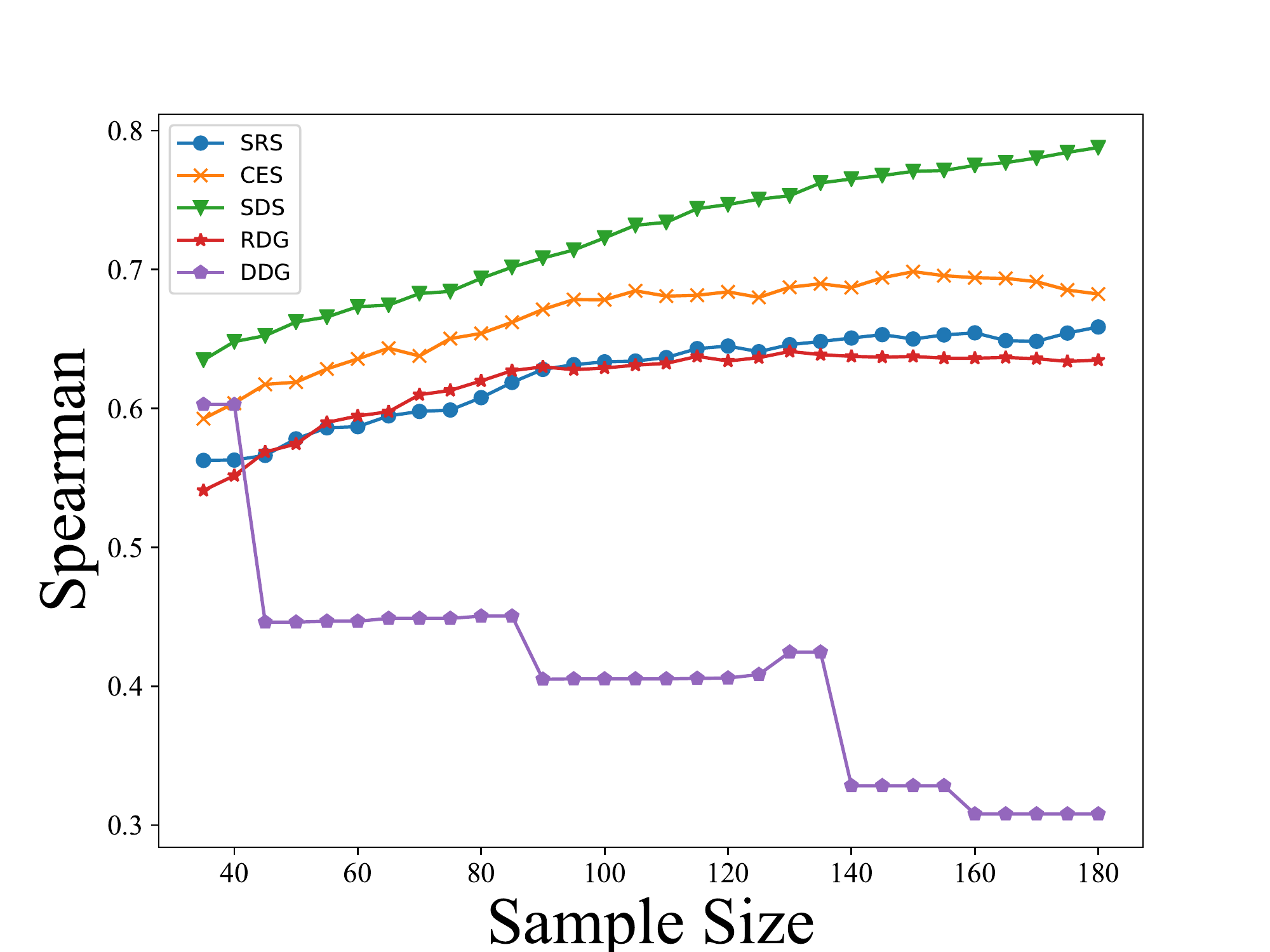}}
	\subfigure[Spearman coefficient of FASHION-MNIST]{
		\label{Main-F-S} 
		\includegraphics[width=0.31\textwidth]{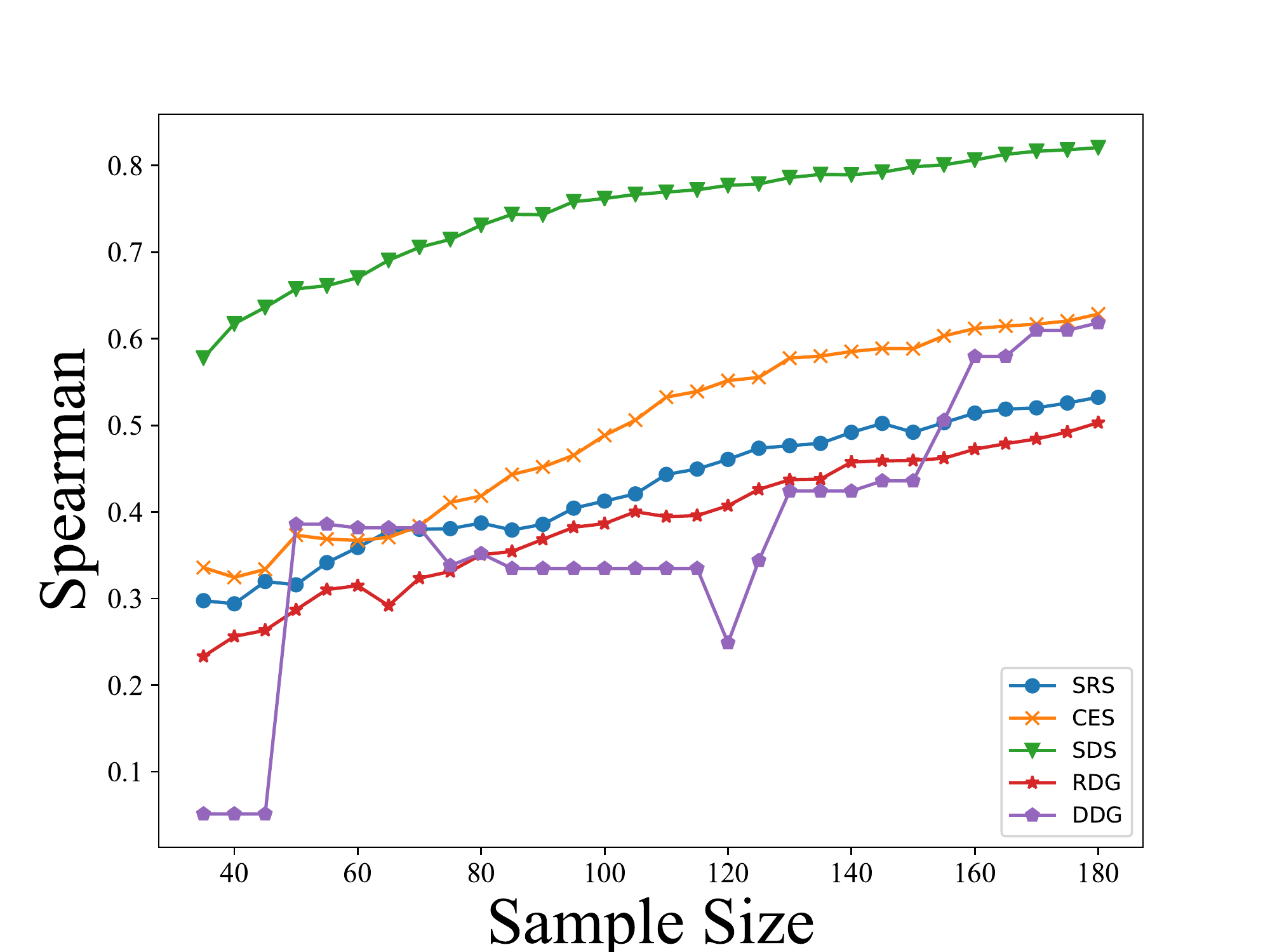}}
	\subfigure[Spearman coefficient of CIFAR-10]{
		\label{Main-C-S} 
		\includegraphics[width=0.31\textwidth]{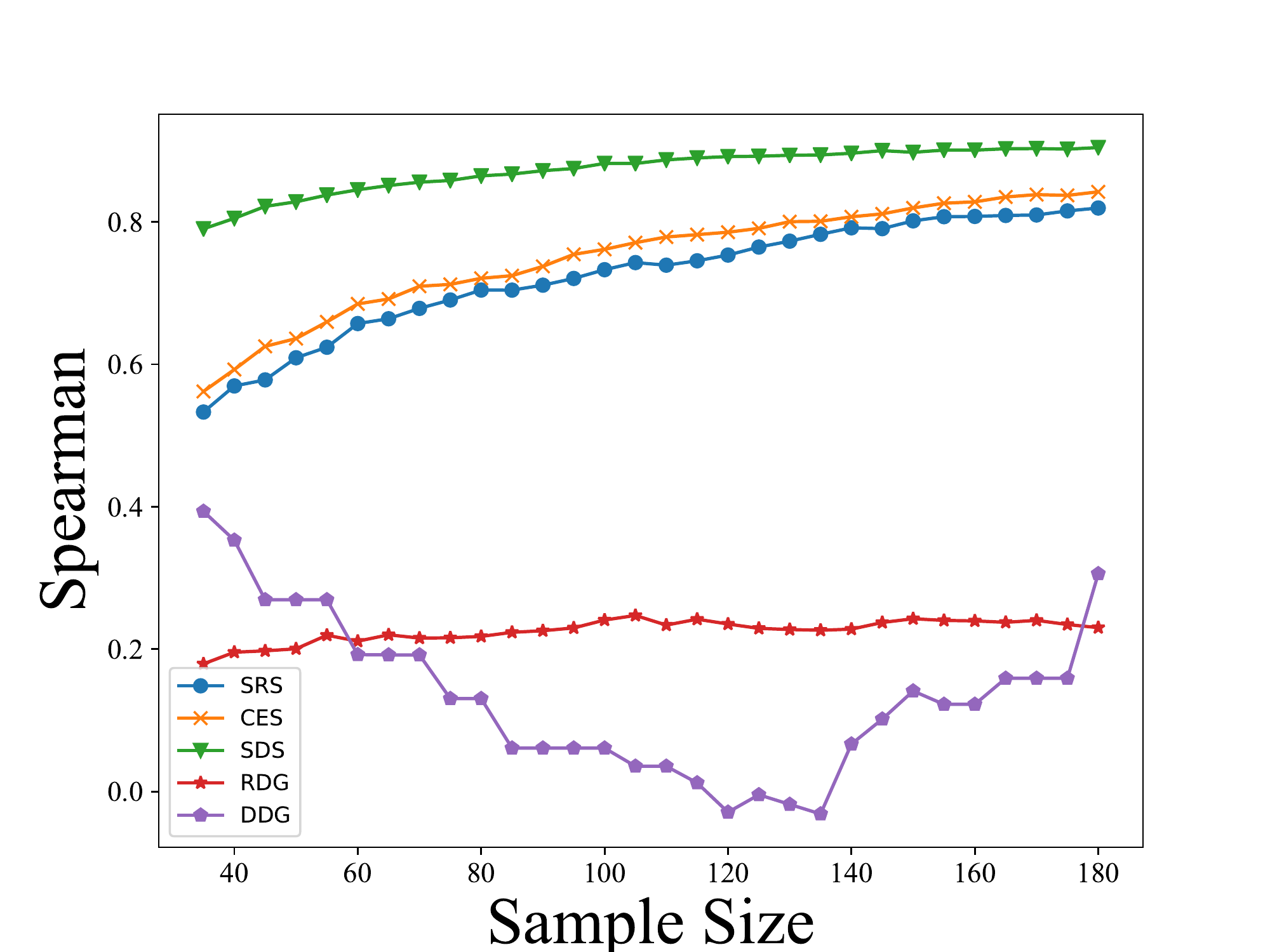}}	
	\subfigure[Jaccard coefficient of MNIST]{
		\label{Main-M-J} 
		\includegraphics[width=0.31\textwidth]{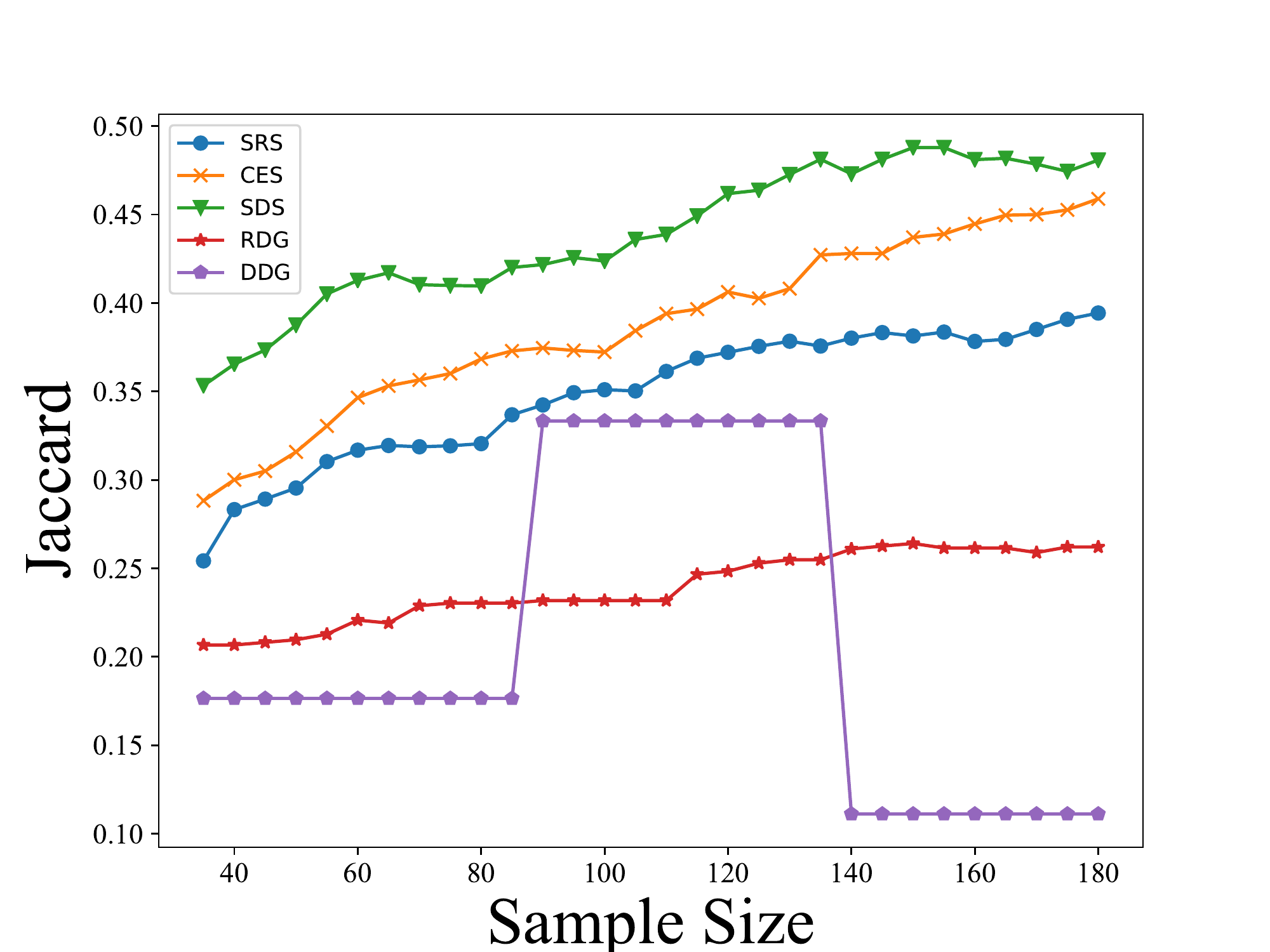}}
	\subfigure[Jaccard coefficient of FASHION-MNIST]{
		\label{Main-F-J} 
		\includegraphics[width=0.31\textwidth]{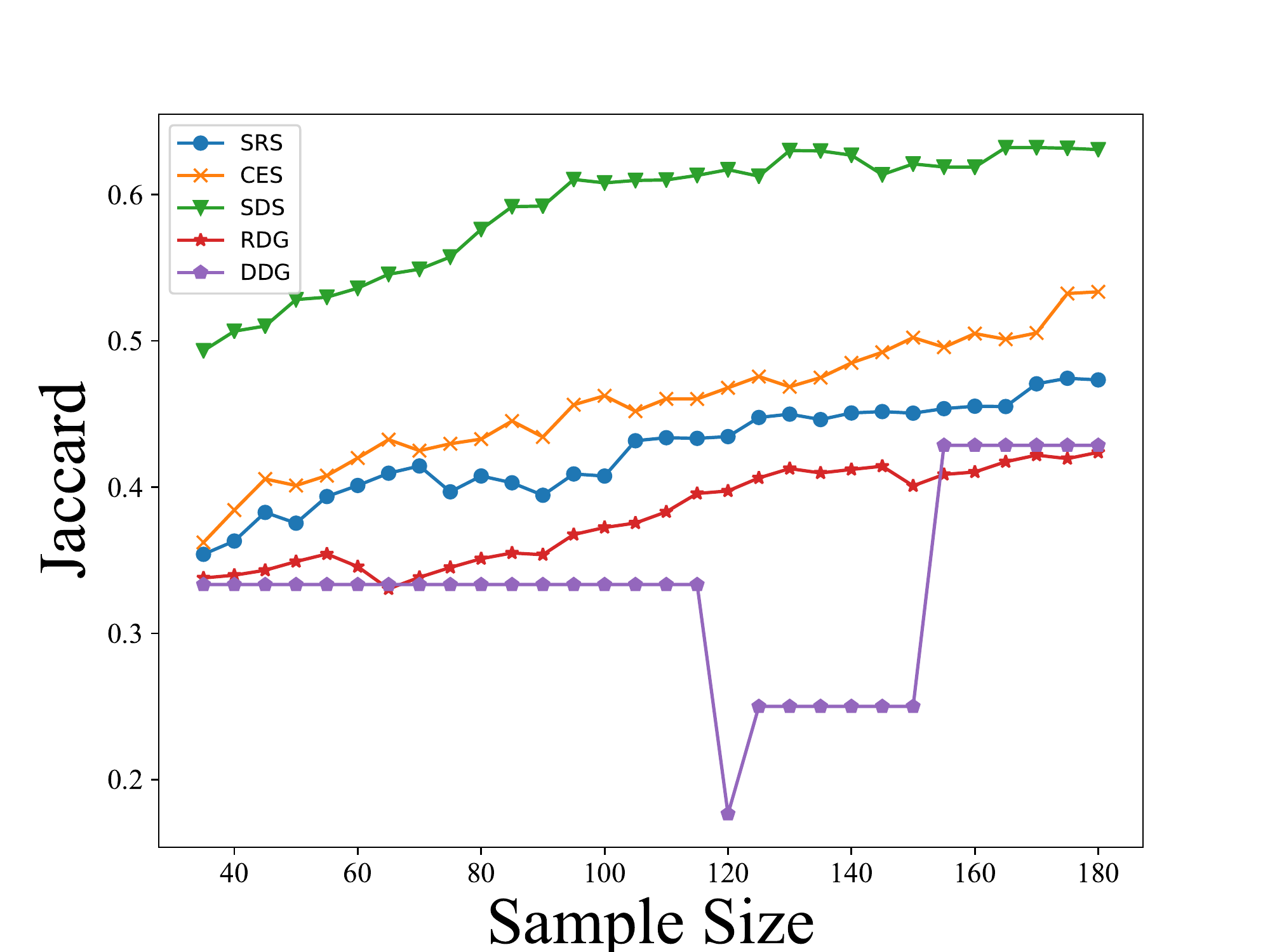}}
	\subfigure[Jaccard coefficient of CIFAR-10]{
		\label{Main-C-J} 
		\includegraphics[width=0.31\textwidth]{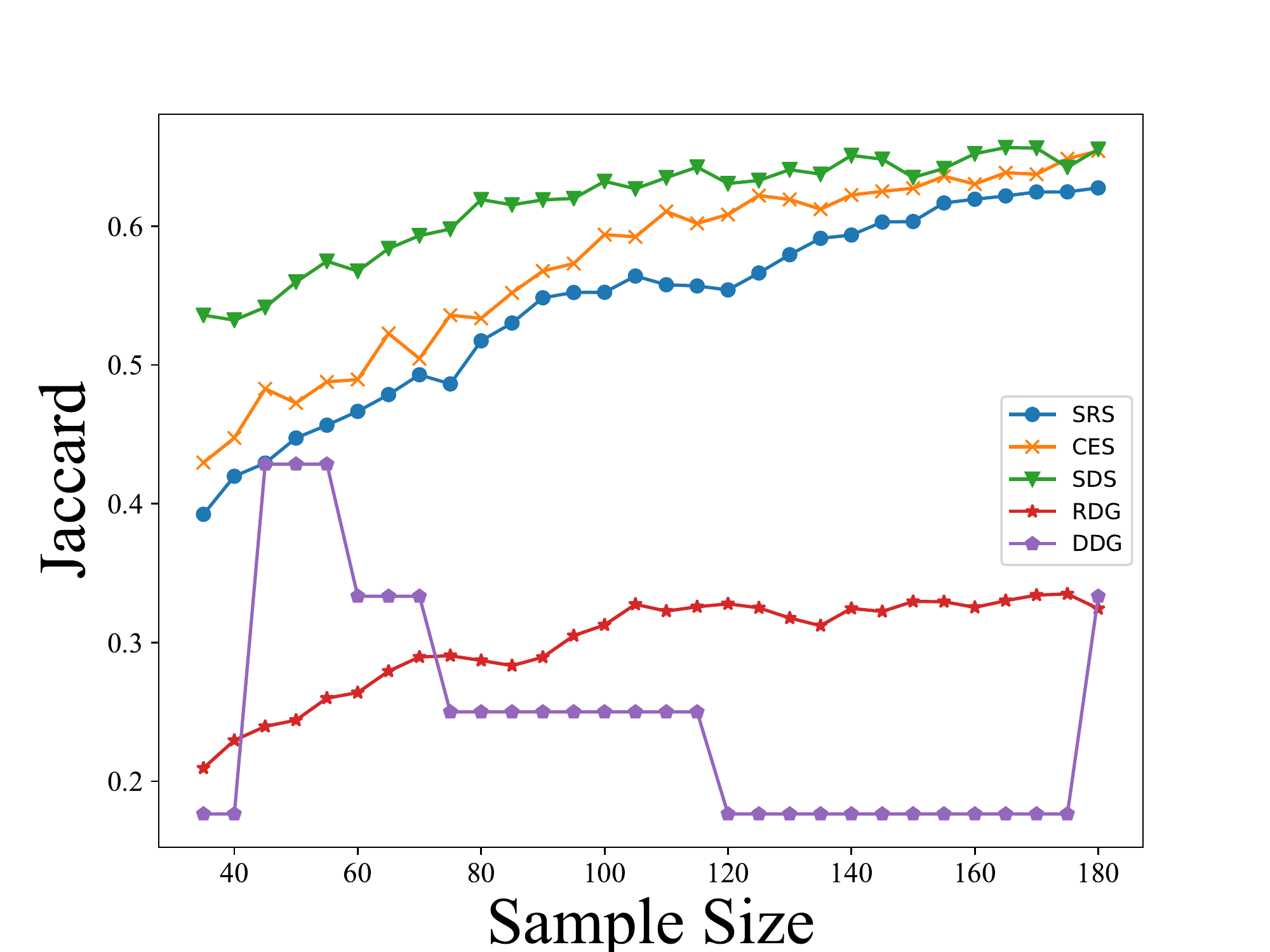}}
	\caption{The results of our approach and the four baselines for ranking model performance under Spearman coefficient and Jaccard coefficient ($k=10$).  
 In each subfigure, the $x$-axis indicates the number of samples, from 35 to 180, and the $y$-axis represents the values of Spearman/Jaccard coefficient. These five studied methods are denoted by lines with different colors, i.e., {\color{green}{$\blacksquare$}} for SDS, {\color{orange}{$\blacksquare$}} for CES, {\color{violet}{$\blacksquare$}} for DDG, {\color{red}{$\blacksquare$}} for RDG, and {\color{blue}{$\blacksquare$}} for SRS.}
	\label{main_ex}
\end{figure*}

\subsection{Evaluation Indicators}
To evaluate to what extend the estimated rank w.r.t. selected samples are identical with the actual rank w.r.t. the whole testing context, we introduce two indicators Spearman's rank correlation coefficient and   Jaccard similarity coefficient.

Spearman's rank correlation coefficient  $\rho$  is a measure of the correlation between two variables $X$ and $Y$ \cite{beller2015and,thongtanunam2020review}. It can be calculated by the following formula: 
\[\rho = \frac{\sum_{i} (x_i-\bar{x})(y_i-\bar{y})}{\sqrt{\sum_{i} (x_i-\bar{x})^2\sum_{i}(y_i-\bar{y})^2}}\]
The value of $\rho$ ranges from -1 to 1: the closer it is to 1 (-1), the two sets of variables are positively (negatively) correlated.

Besides, we introduce Jaccard similarity coefficient \cite{tian2018deeptest,zhou2015metamorphic} (denoted as $J_k$) to evaluate the similarity between the top-$k$ model sets generated by the estimated rank and actual rank. For example, the estimated rank is $\mathcal{M}_1,\mathcal{M}_3,\mathcal{M}_5,\cdots $, and the actual rank is $\mathcal{M}_1,\mathcal{M}_3,\mathcal{M}_2,\cdots$. The Jaccard similarity coefficient between two top-3 model sets $\{\mathcal{M}_1,\mathcal{M}_3,\mathcal{M}_5\}$ and $\{\mathcal{M}_1,\mathcal{M}_2,\mathcal{M}_3\}$ is calculated as:  
\[J_3=\frac{|\{\mathcal{M}_1,\mathcal{M}_3,\mathcal{M}_5\}\cap\{\mathcal{M}_1,\mathcal{M}_2,\mathcal{M}_3\}|}{|\{\mathcal{M}_1,\mathcal{M}_3,\mathcal{M}_5\}\cup\{\mathcal{M}_1,\mathcal{M}_2,\mathcal{M}_3\}|}=\frac{2}{4}=0.5\]  
As we encounter dozens of models in the testing context (e.g., 28 models for MNIST), we focus on $k=1,3,5,10$ to evaluate the performance of our method on different cutoff points. We take $k=10$ as the representative  to report the evaluation under Jaccard similarity coefficient in the experimental results. The other Jaccard coefficient (when $k=1, 3, 5$) will be discussed in the discussion part.

\subsection{Analysis Method}

First, we employ Wilcoxon rank sum test \cite{rank_sum} to verify the difference of the rank performance between our method and the baselines. If the $p$-value are less than 0.05, the two sets of data are considered significantly different.

Next, we introduce Cliff's delta $\delta$ \cite{cliff_delta}, which measures the effect size for comparing two ordinal data lists. We judge the difference between the two sets of data based on the range of $\delta$: negligible, if $|\delta|<0.147$; small, if $0.147\leq |\delta|<0.330$; medium, if $0.330\leq |\delta|<0.474$, and large, if $|\delta|\geq 0.474$.

Finally, we use ``W/T/L'' \cite{WTL,liu2018connecting} to compare the results of our approach and the baseline, where ``W'' means our approach wins, ``T'' means the results are tie, and ``L'' means our approach loses. Reaching the two standards shows that our approach wins: (a) the $p$-value of Wilcoxon rank sum test is less than 0.05 ($p<0.05$), which means the results between our approach and baseline are significantly different; (b) the Cliff's delta $\delta$ is larger than 0.147 ($\delta>0.147$), which means the difference between the two results are positive and not negligible. If $p<0.05$ and $\delta<-0.147$,  we consider our approach loses. Otherwise, the result of comparision is tie.

\subsection{Research Questions}

We are committed to promoting the ranking performance of the multiple models under limited labeling effort in the comparative testing scenario. We propose the following two research questions (RQs) to organize our experiments:

\begin{itemize}
	\item RQ1 (Effectiveness): Whether our method SDS can surpass the state-of-the-art methods in ranking multiple models?
	\item RQ2 (Efficiency): Compared with  the state-of-the-art methods, is our method SDS efficient?
\end{itemize}

\section{EXPERIMENT RESULTS}
\label{sec:expres}
In this section, we present the results of the experiments and answer the above two RQs.

\subsection{RQ1: Effectiveness}
\textbf{Motivation and Approach.} Our problem is to obtain effective ranking results of model performance with a very low labeling percent of testing samples for multiple models in the testing scenario. We hope to verify whether our proposed approach SDS is more effective than the baseline methods with limited labelling effort. To achieve this aim, we compare the ranking performance of SDS with the baseline methods in three testing contexts (containing the 10000 testing samples from MNIST, Fashion-MNIST, and CIFAR-10, respectively) under the sampling sizes from 35 to 180 with intervals of 5 (i.e., 35, 40, \dots, 180). Specifically, we employ these five studied methods to sample the subset under different sampling sizes, and use the ranking result on the subset to estimate the rank on the whole testing context. We repeat running these methods 50 times, and report the average of calculated results.

\textbf{Results.} Figure~\ref{main_ex} shows the comparison results of our approach and the four baselines for ranking model performance. The three subgraphs in the first row show the comparison results of Spearman coefficient $\rho$, and the subgraphs in the second row present the results of Jaccard coefficient\footnote{We take $k=10$ as the representative to report the evaluation under Jaccard similarity coefficient in the experimental results. The other Jaccard coefficient (when $k=1, 3, 5$) will be discussed in the discussion part.} ($J_{10}$). In each subfigure, the $x$-axis indicates the number of samples, from 35 to 180, and the $y$-axis represents the values of Spearman/Jaccard coefficient. These five studied methods are denoted by lines with different colors, i.e., {\color{green}{$\blacksquare$}} for SDS, {\color{orange}{$\blacksquare$}} for CES, {\color{violet}{$\blacksquare$}} for DDG, {\color{red}{$\blacksquare$}} for RDG, and {\color{blue}{$\blacksquare$}} for SRS. 
 It can be seen in Figure~\ref{main_ex} that in all sub-graphs, our method SDS is obviously better than the other baselines under all sampling sizes from 35 to 180. Besides, our approach is very stable; on the contrary, some baselines have a strong volatility, e.g., DDG has wild gyrations when measuring Jaccard coefficient for MINIST, Fashion-MINIST, and CIFAR-10.

\begin{table*}
	\caption{The results of Spearman correlation and Jaccard correlation ($J_{10}$) with our method and four baseline methods. }
	\label{tab:main}
	\renewcommand\arraystretch{1.1} 
	\centering 
	\scalebox{0.85}{ 
		\begin{tabular}{|p{1.2cm}|p{1.0cm}|p{0.65cm}p{0.65cm}p{0.65cm}p{0.65cm}p{0.65cm}|p{0.65cm}p{0.65cm}p{0.65cm}p{0.65cm}p{0.65cm}|p{0.65cm}p{0.65cm}p{0.65cm}p{0.8cm}p{0.65cm}|}
			\hline
			\multirow{2}*{Indicator} &  \multirow{2}*{Cutoff}       & \multicolumn{5}{c|}{MNIST}                                                   & \multicolumn{5}{c|}{FASHION-MNIST}                                           & \multicolumn{5}{c|}{CIFAR-10}                                                   \\ 
			&       & SRS & CES & RDG & DDG & SDS             & SRS & CES & RDG & DDG & SDS             & SRS & CES & RDG & DDG & SDS             \\ 
			\hline
			\hline
			\multirow{6}{*}{Spearman} &35  &\cellcolor{gray} 0.563  & 0.593        & 0.541\cellcolor{gray}     &\cellcolor{gray} 0.603         & \textbf{0.635} &\cellcolor{gray} 0.298  &\cellcolor{gray} 0.336        &\cellcolor{gray} 0.233   &\cellcolor{gray} 0.051         & \textbf{0.578} &\cellcolor{gray} 0.533  &\cellcolor{gray} 0.562      &\cellcolor{gray} 0.179     &\cellcolor{gray} 0.393         & \textbf{0.790} \\
			& 60  &\cellcolor{gray} 0.587  &0.636        &\cellcolor{gray} 0.595     &\cellcolor{gray} 0.447       & \textbf{0.673} & \cellcolor{gray}0.359  &\cellcolor{gray} 0.367        & \cellcolor{gray}0.315  & \cellcolor{gray}0.382         & \textbf{0.670} &\cellcolor{gray} 0.657  & \cellcolor{gray}0.685     &\cellcolor{gray} 0.211     &\cellcolor{gray} 0.192         & \textbf{0.845} \\
			& 90 &\cellcolor{gray} 0.628  & 0.671        &\cellcolor{gray} 0.630     &\cellcolor{gray} 0.405         & \textbf{0.708} &\cellcolor{gray} 0.386  &\cellcolor{gray} 0.452        &\cellcolor{gray} 0.368     &\cellcolor{gray} 0.335         & \textbf{0.743} &\cellcolor{gray} 0.711  &\cellcolor{gray} 0.738    &\cellcolor{gray} 0.226     &\cellcolor{gray} 0.061         & \textbf{0.872} \\
			& 120 &\cellcolor{gray} 0.645  &\cellcolor{gray} 0.684        &\cellcolor{gray} 0.634     &\cellcolor{gray} 0.405       & \textbf{0.747} &\cellcolor{gray} 0.461  &\cellcolor{gray} 0.552        &\cellcolor{gray} 0.407    &\cellcolor{gray} 0.249         & \textbf{0.777} &\cellcolor{gray} 0.753  &\cellcolor{gray} 0.785      &\cellcolor{gray} 0.235     &\cellcolor{gray} -0.029        & \textbf{0.892} \\
			& 150 &\cellcolor{gray} 0.650  &\cellcolor{gray} 0.699        &\cellcolor{gray} 0.637     &\cellcolor{gray} 0.328         & \textbf{0.771} & \cellcolor{gray}0.492  &\cellcolor{gray} 0.589        &\cellcolor{gray} 0.460   & \cellcolor{gray}0.436         & \textbf{0.798} &\cellcolor{gray} 0.801  &\cellcolor{gray} 0.820     &\cellcolor{gray} 0.243     &\cellcolor{gray} 0.141         & \textbf{0.898} \\
			& 180 &\cellcolor{gray} 0.659  &\cellcolor{gray} 0.682        &\cellcolor{gray} 0.635     &\cellcolor{gray} 0.308         & \textbf{0.788} & \cellcolor{gray}0.532  &\cellcolor{gray} 0.629        & \cellcolor{gray}0.503     &\cellcolor{gray} 0.618        & \textbf{0.821} &\cellcolor{gray} 0.819  &\cellcolor{gray} 0.842      &\cellcolor{gray} 0.230     &\cellcolor{gray} 0.306         & \textbf{0.904} \\ \hline 
			& Average &0.622 &0.661 &0.612 &0.416 &\textbf{0.720} &0.421 &0.487 &0.381 &0.345 &\textbf{0.731}	&0.713	&0.739	&0.221	&0.177	&\textbf{0.867}\\ 
			&W/T/L	&6/0/0	&3/3/0	&6/0/0	&6/0/0	&/	&6/0/0	&6/0/0	&6/0/0	&6/0/0	&/
			&6/0/0	&6/0/0	&6/0/0	&6/0/0 	&/\\ 
			\hline
			\hline
			\multirow{6}{*}{Jaccard}  & 35  &\cellcolor{gray} 0.254  &\cellcolor{gray} 0.288        &\cellcolor{gray} 0.207     & \cellcolor{gray}0.177         & \textbf{0.353} &\cellcolor{gray} 0.354  &\cellcolor{gray} 0.362        &\cellcolor{gray} 0.338    &\cellcolor{gray} 0.333         & \textbf{0.493} &\cellcolor{gray} 0.392  & \cellcolor{gray}0.430        &\cellcolor{gray} 0.210     &\cellcolor{gray} 0.177         & \textbf{0.536} \\
			& 60  &\cellcolor{gray} 0.317  & \cellcolor{gray}0.347        & \cellcolor{gray}0.221     &\cellcolor{gray} 0.177        & \textbf{0.413} & \cellcolor{gray}0.401  & \cellcolor{gray}0.420        &\cellcolor{gray} 0.346    &\cellcolor{gray} 0.333         & \textbf{0.536} &\cellcolor{gray} 0.467  &\cellcolor{gray} 0.490       &\cellcolor{gray} 0.264     &\cellcolor{gray} 0.333         & \textbf{0.568} \\
			& 90  &\cellcolor{gray} 0.342  & 0.375        &\cellcolor{gray} 0.232     &\cellcolor{gray} 0.333        & \textbf{0.422} & \cellcolor{gray}0.394  &\cellcolor{gray} 0.434        &\cellcolor{gray} 0.354    &\cellcolor{gray} 0.333         & \textbf{0.592} &\cellcolor{gray} 0.548  & 0.568       &\cellcolor{gray} 0.289     &\cellcolor{gray} 0.250         & \textbf{0.619} \\
			& 120 &\cellcolor{gray} 0.372  &\cellcolor{gray} 0.406        &\cellcolor{gray} 0.248     &\cellcolor{gray} 0.333        & \textbf{0.462} & \cellcolor{gray}0.435  & \cellcolor{gray}0.468        &\cellcolor{gray} 0.397     &\cellcolor{gray} 0.177         & \textbf{0.617} &\cellcolor{gray} 0.554  & 0.608       & \cellcolor{gray}0.328     &\cellcolor{gray} 0.177         & \textbf{0.631} \\
			& 150 &\cellcolor{gray} 0.381  &\cellcolor{gray} 0.437       &\cellcolor{gray} 0.264     &\cellcolor{gray} 0.111        & \textbf{0.488} & \cellcolor{gray}0.451  &\cellcolor{gray} 0.502        &\cellcolor{gray} 0.401     & \cellcolor{gray}0.250         & \textbf{0.621} & 0.603  & 0.627       & \cellcolor{gray}0.330     &\cellcolor{gray} 0.177         & \textbf{0.635} \\ 
			& 180 &\cellcolor{gray} 0.394  & 0.459        &\cellcolor{gray} 0.262     &\cellcolor{gray} 0.111                  & \textbf{0.481} &\cellcolor{gray} 0.473  &\cellcolor{gray} 0.534        & \cellcolor{gray}0.424              &\cellcolor{gray} 0.429         & \textbf{0.631} & 0.628  & 0.654                 & \cellcolor{gray}0.324     & \cellcolor{gray}0.333         & \textbf{0.656} \\
			\hline
			& Average &0.344 	&0.385 	&0.239 	&0.207 	&\textbf{0.436} 	&0.418 	&0.453 	&0.377 	&0.309 	&\textbf{0.582} 	&0.532 	&0.563 	&0.291 	&0.241 	&\textbf{0.607} 
			\\ 
			&W/T/L	&6/0/0	&4/2/0	&6/0/0	&6/0/0	&/	&6/0/0	&6/0/0	&6/0/0	&6/0/0	&/
			&4/2/0	&2/4/0	&6/0/0	&6/0/0 	&/\\ \hline
		\end{tabular}
	}
\end{table*}

In order to show more details of the experiment results, we choose six sampling points (35, 60, 90, 120, 150, and 180) as the representatives. Table~\ref{tab:main} presents the detailed results under these six points, with the mean values of Spearman coefficient and Jaccard coefficient of ranking multiple models obtained by 50 repetitions of running the five studied methods. The best numbers are highlighted in bold. In Table~\ref{tab:main}, if our approach wins the baseline method (that is to say, the $p$ value is less than 0.05 and the $\delta$ is greater than 0.147), then we add the gray background to the value of the baseline method. Based on the values of Spearman and Jaccard coefficients, we have added two rows: ``Average'' to calculate the average value of each column and ``W/T/L'' to record the number of times our approach win/tie/lose other baselines.

From Table~\ref{tab:main}, we have the following observations. (a) From the average value of each point, our approach is higher than all baselines, under both Spearman coefficient and Jaccard coefficient. (b) The gray background indicates that our approach wins other baselines at the most of points.
 (c) The results of W/T/L shows that our approach is not only higher than other baselines in mean, but also significantly better.

\begin{framed}
\noindent Answer to RQ1: In ranking multiple DL models, 
our approach is significantly better than all other baselines in effectiveness.
\end{framed}
\subsection{RQ2: Efficiency}
\textbf{Motivation and Approach.} In RQ1, we have observed that our approach SDS is significantly better than other baselines under both Spearman coefficient and Jaccard coefficient in raking multiple DL models. 
The process of sample selection may be time consuming. In this RQ, we want to check the efficiency of our approach compared with other baselines.

\textbf{Results.} 
Table~\ref{tab:eff} shows the total time consumed when running studied methods with sampling from 35 to 180. 
From Table 3, we find that our approach SDS takes longer than SRS because it contains sample sorting and operations on the prediction matrix. The time SDS consumes is similar to the other three baselines CES, RDG, and DDG, which is around 10,000 seconds.

\begin{table}[t]
	\caption{The time (second) consumed when samples are selected by different approaches.}
	\label{tab:eff}
	\renewcommand\arraystretch{1.2} 
	\centering
	\begin{tabular}{crrrrr}
		
		\hline
		dataset       & SRS   & CES & RDG   & DDG & SDS   \\
		\hline
		MNIST         & 1,117 & 3,513  & 11,334 & 9,493  & 10,256 \\
		Fashion-MNIST & 1,403 & 31,703  & 10,082 & 8,538  & 9,179 \\
		CIFAR-10       & 4,799 & 15,347  & 11,679 & 9,669  & 10,642\\
		\hline
	\end{tabular}
\end{table}
\begin{framed}
	\noindent Answer to RQ2: Except for SRS, our approach is similar to other baselines in time consumption.
\end{framed}

\section{Discussion}
\label{sec:dis}
In this section, we further discuss some parameter settings and results in the experiments. First, we analyze the parameter and indicator involved in the experiments. After that, we discuss why our algorithm can effectively help multi-model performance ranking and whether our method is effective when the number of models is reduced. 

\subsection{The performance under other selection rates}
\begin{figure}[]
	\centering
	\subfigure[Spearman of MNIST]{
		\label{DIS1-M-S} 
		\includegraphics[width=0.23\textwidth]{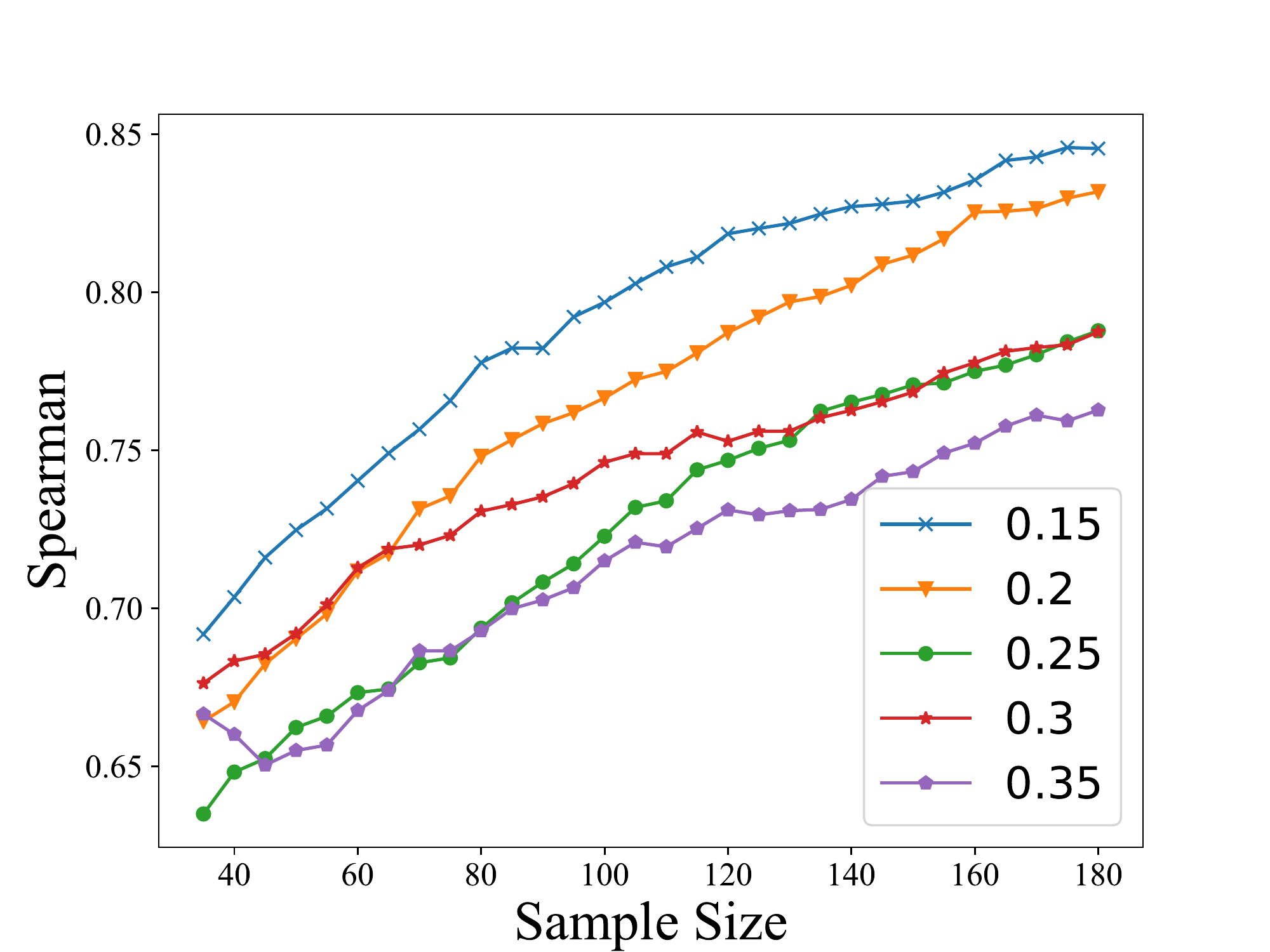}}
	\subfigure[Jaccard of MNIST]{
		\label{DIS1-M-J} 
		\includegraphics[width=0.23\textwidth]{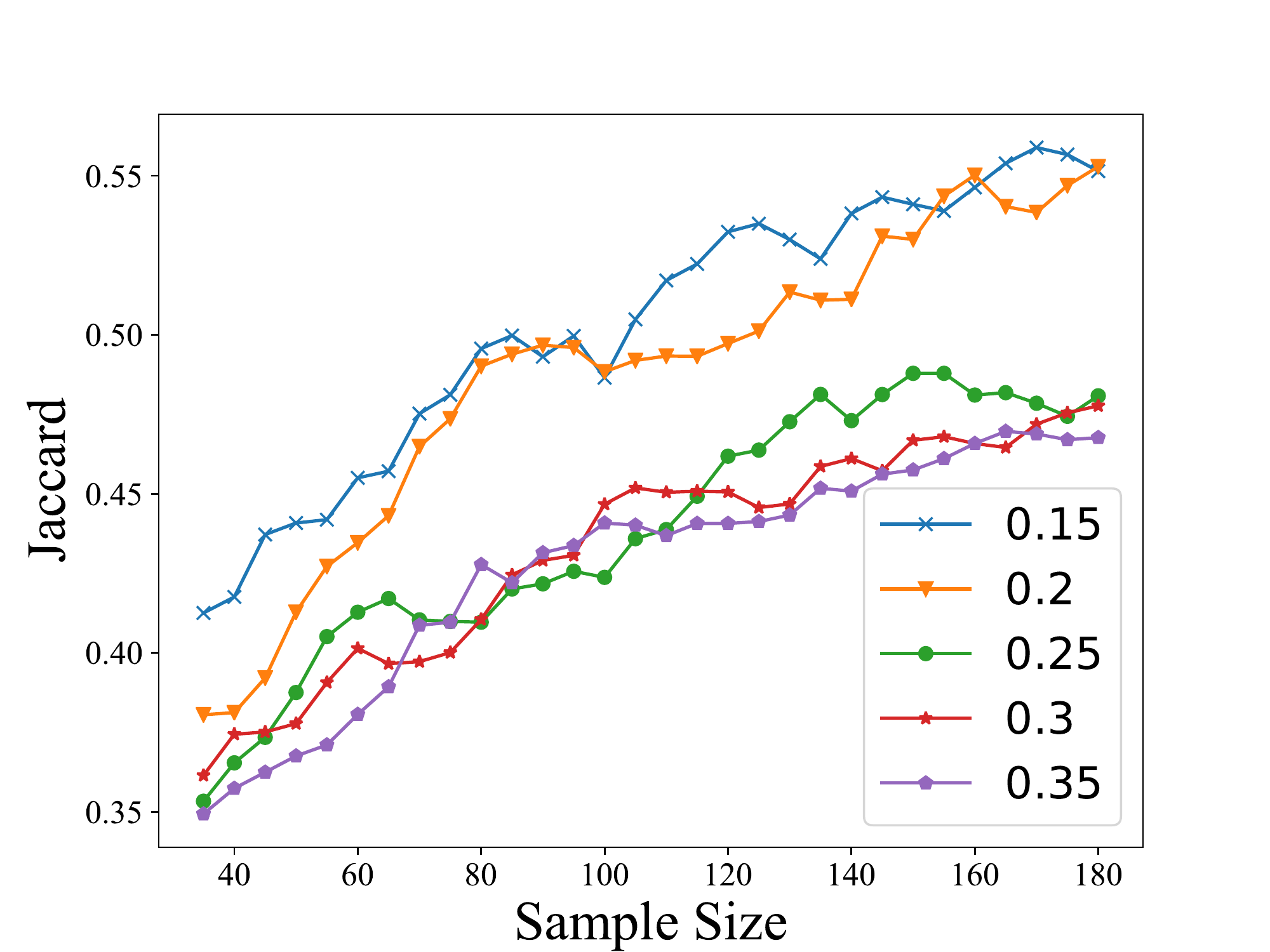}}
	\subfigure[Spearman of FASHION-MNIST]{
		\label{DIS1-F-S} 
		\includegraphics[width=0.23\textwidth]{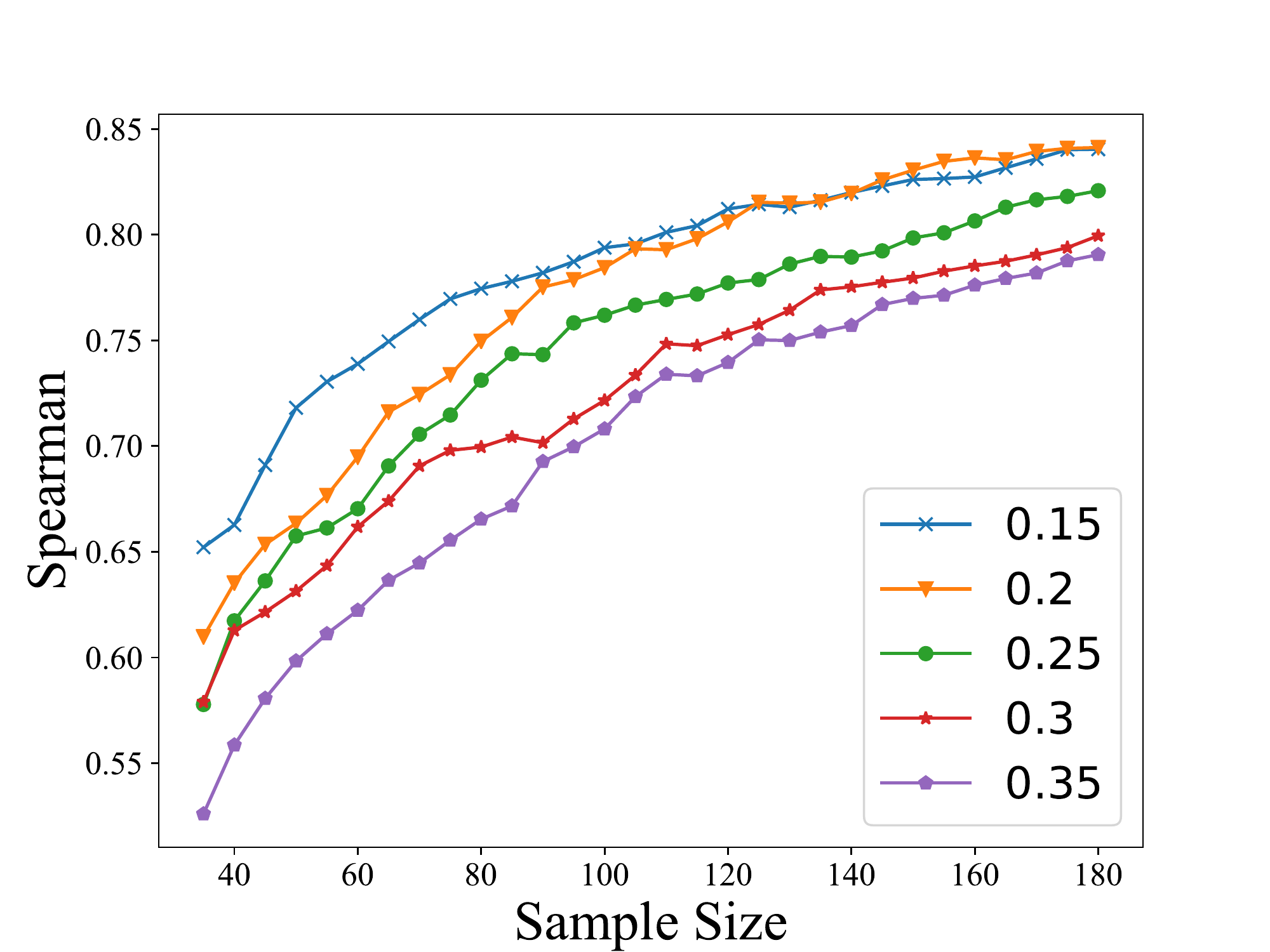}}
	\subfigure[Jaccard of FASHION-MNIST]{
		\label{DIS1-F-J} 
		\includegraphics[width=0.23\textwidth]{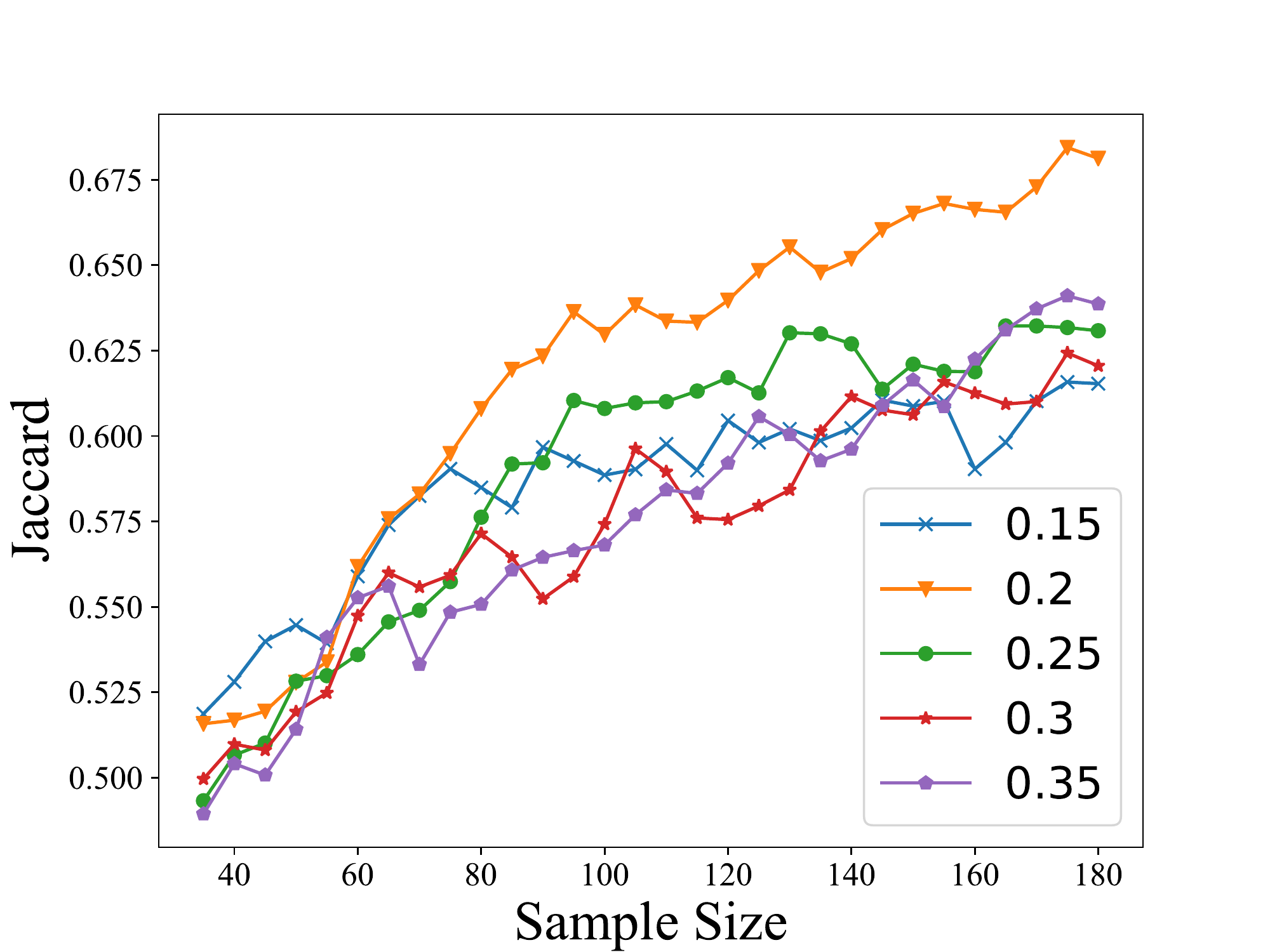}}
	\subfigure[Spearman of CIFAR-10]{
		\label{DIS1-C-S} 
		\includegraphics[width=0.23\textwidth]{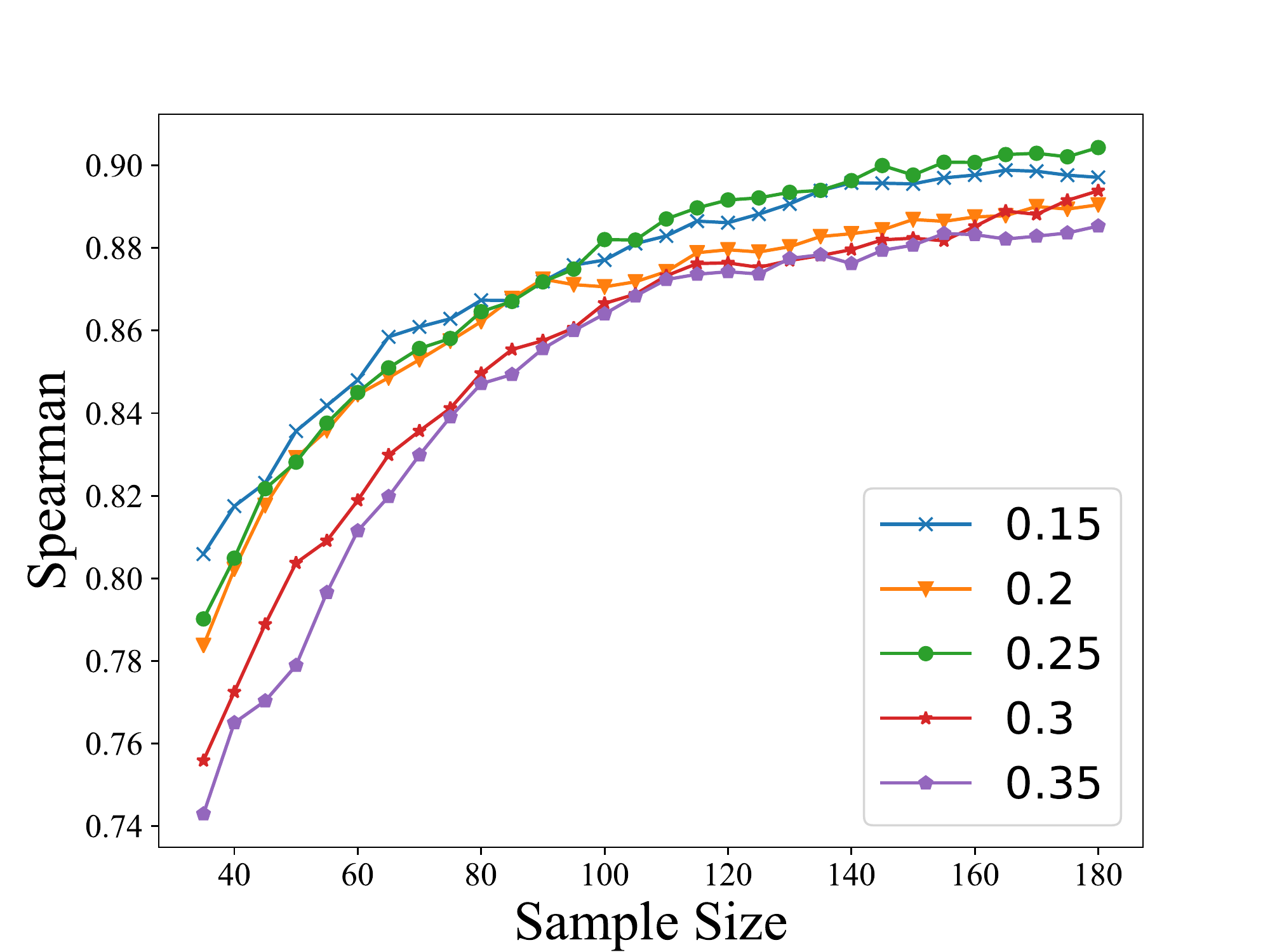}}	
	\subfigure[Jaccard of CIFAR-10]{
		\label{DIS1-C-J} 
		\includegraphics[width=0.23\textwidth]{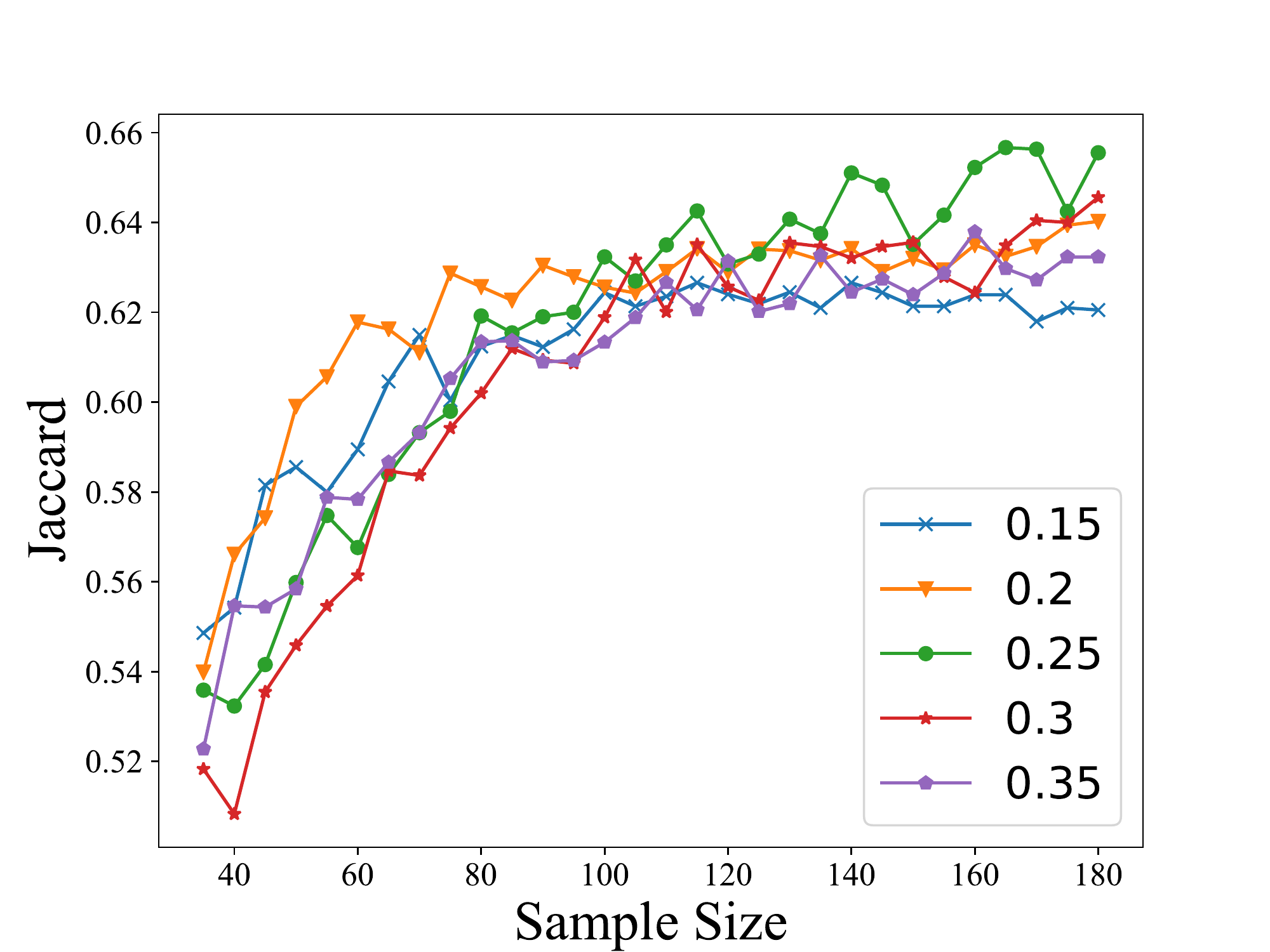}}
	\caption{The graph of ranking performance when the random selection rates changes from top 15\% - top 35\%. The performances of different rates are denoted by line with different colors, i.e., {\color{blue}{$\blacksquare$}} for 15\%, {\color{orange}{$\blacksquare$}} for 20\%, {\color{green}{$\blacksquare$}} for 25\%, {\color{red}{$\blacksquare$}} for 30\%, and {\color{violet}{$\blacksquare$}} for 35\%.}
	\label{dis_1}
\end{figure}

In our experiment, the random sampling interval is set to the top 25\% as shown in Step 5 of Algorithm 1. We want to further discuss the performance under other selection rates by conducting experiments on five different selection rates (i.e., random sampling of the top 15\%, 20\%, 25\%, 30\%, and 35\% intervals).  The results are shown in Figure~\ref{dis_1}, where the performances of different rates are denoted by line with different colors, i.e., {\color{blue}{$\blacksquare$}} for 15\%, {\color{orange}{$\blacksquare$}} for 20\%, {\color{green}{$\blacksquare$}} for 25\%, {\color{red}{$\blacksquare$}} for 30\%, and  {\color{violet}{$\blacksquare$}} for 35\%, respectively.      

It can be seen that the performances of different selection rates on different datasets vary a lot. Generally speaking, there is no obvious trend in all subfigures. In addition, the 25\% sampling interval (the green line) we set in the experiment obtains the best ranking performance under the most of sampling sizes in the CIFAR-10 dataset. As quartering is common for dataset partition in software engineering and easy to implement\cite{Surprise_Adequacy}, we still suggest applying the 25\% interval in our algorithm.

\subsection{The performance of Jaccard coefficient with $k=1,3,5$}
\begin{figure}[]
	\centering
	\subfigure[Jaccard coefficient $J_1$ ($k=1$)]{
		\label{DIS2-J1} 
		\includegraphics[width=0.23\textwidth]{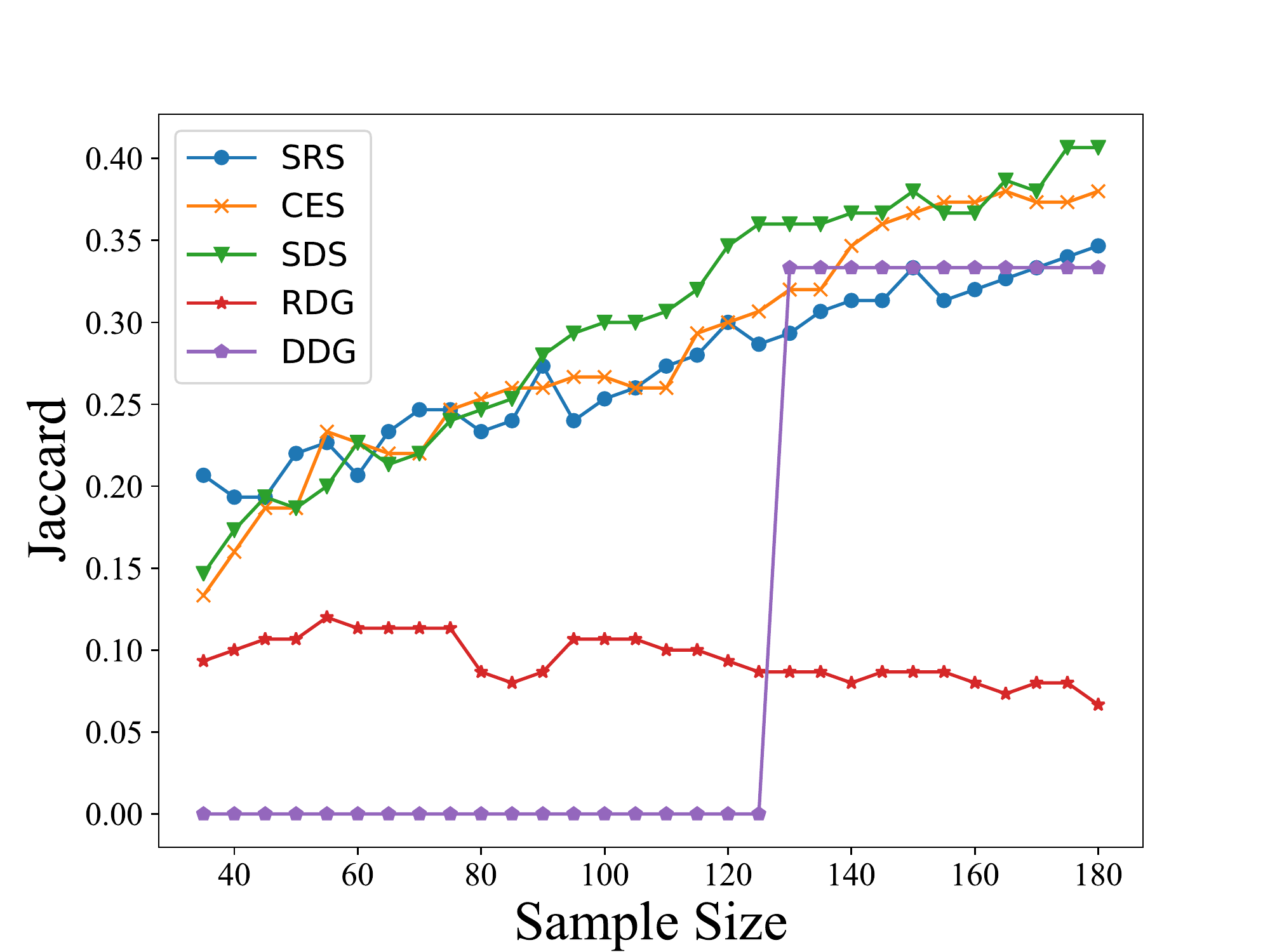}}	
	\subfigure[Jaccard coefficient $J_3$ ($k=3$)]{
		\label{DIS2-J3} 
		\includegraphics[width=0.23\textwidth]{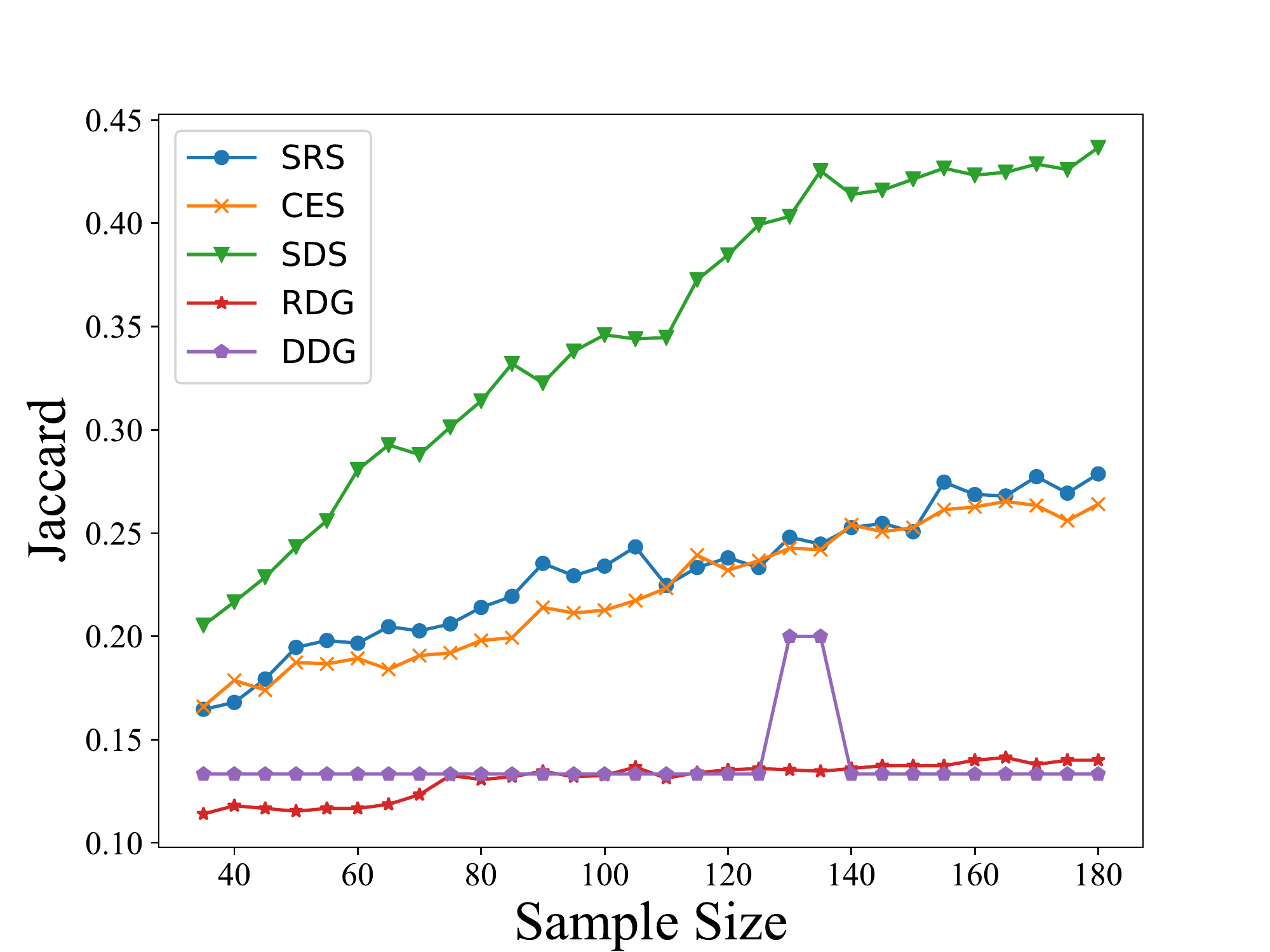}}
	\subfigure[Jaccard coefficient $J_5$ ($k=5$)]{
		\label{DIS2-J5} 
		\includegraphics[width=0.23\textwidth]{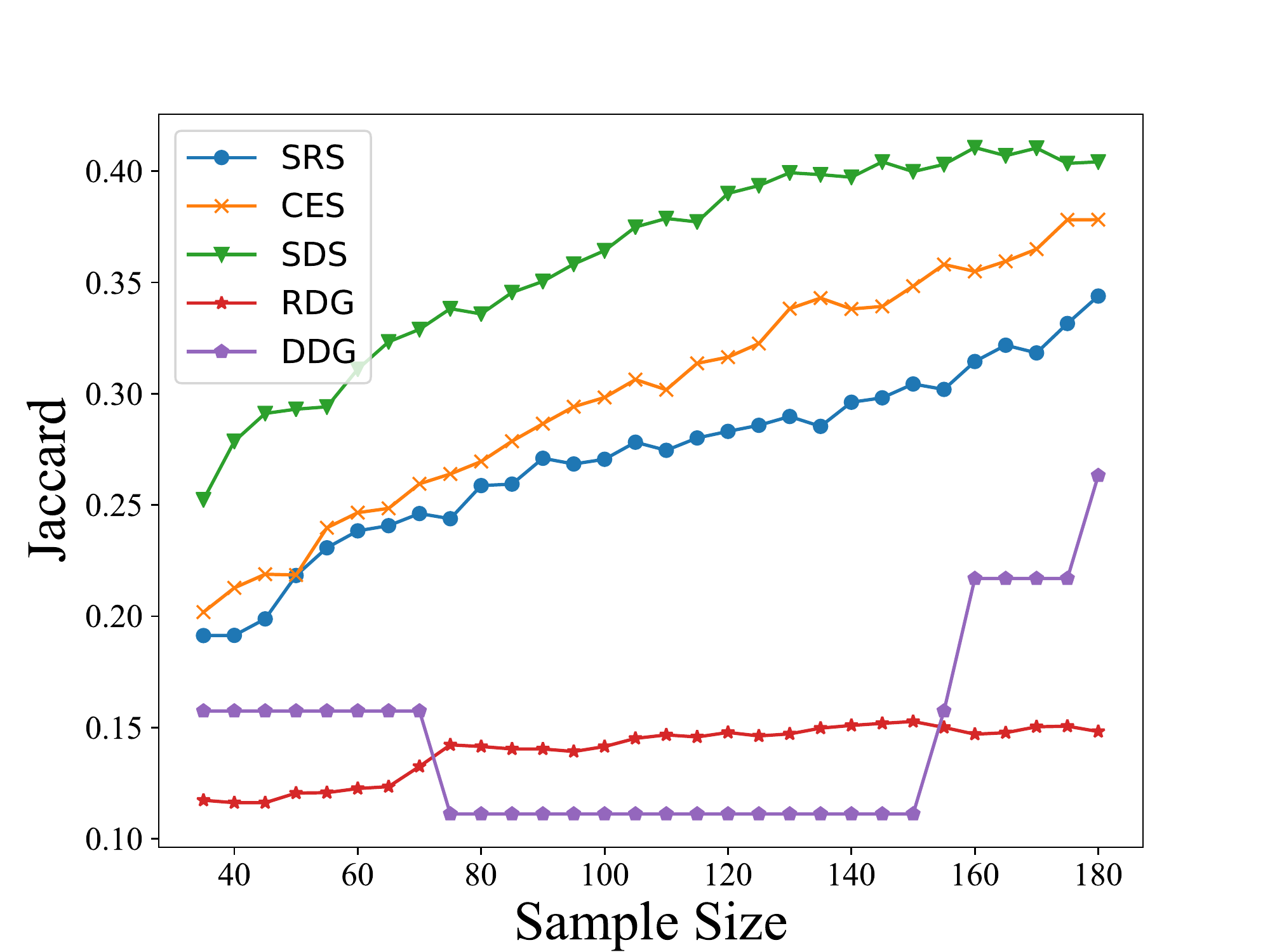}}
	\caption{The graphs for $k=1,3,5$ in measuring Jaccard coefficient with the top-$k$ model sets generated by the selected subset and the whole testing context.}
	\label{dis_2}
\end{figure}

We employ Jaccard coefficient to measure the similarity between the two top-$k$ model sets generated by the selected subset and the whole testing context, respectively. In the previous experiments, 
when we use the Jaccard coefficient, we calculate it with $k=10$. In this section, we will discuss whether our method has advantages when the values of $k$ are different, i.e., $k=1,3,5$. Due to space limitation, we cannot display all the 3$\times$3 (the former 3 for the three datasets and the latter 3 for $k=1,3,5$) subgraphs, we calculate the average of the three datasets in three subgraphs for $k=1,3,5$ in Figure~\ref{dis_2}.

As shown in Figure~\ref{dis_2}, we compare our approach SDS (the green line) with other baselines when $k=1, 3 , 5$. 
Figure~\ref{dis_2} shows the average values of the Jaccard coefficient of the three datasets when the sampling changes. It can be seen that our approach still has advantages under the most of points, which shows that our approach is still superior in ranking models when considering $k=1, 3 , 5$.

\subsection{Analysis and Insight of our algorithm}

\begin{figure}[t]
	\centering
	\subfigure[MNIST]{
		\label{DIS3-M} 
		\includegraphics[width=0.15\textwidth]{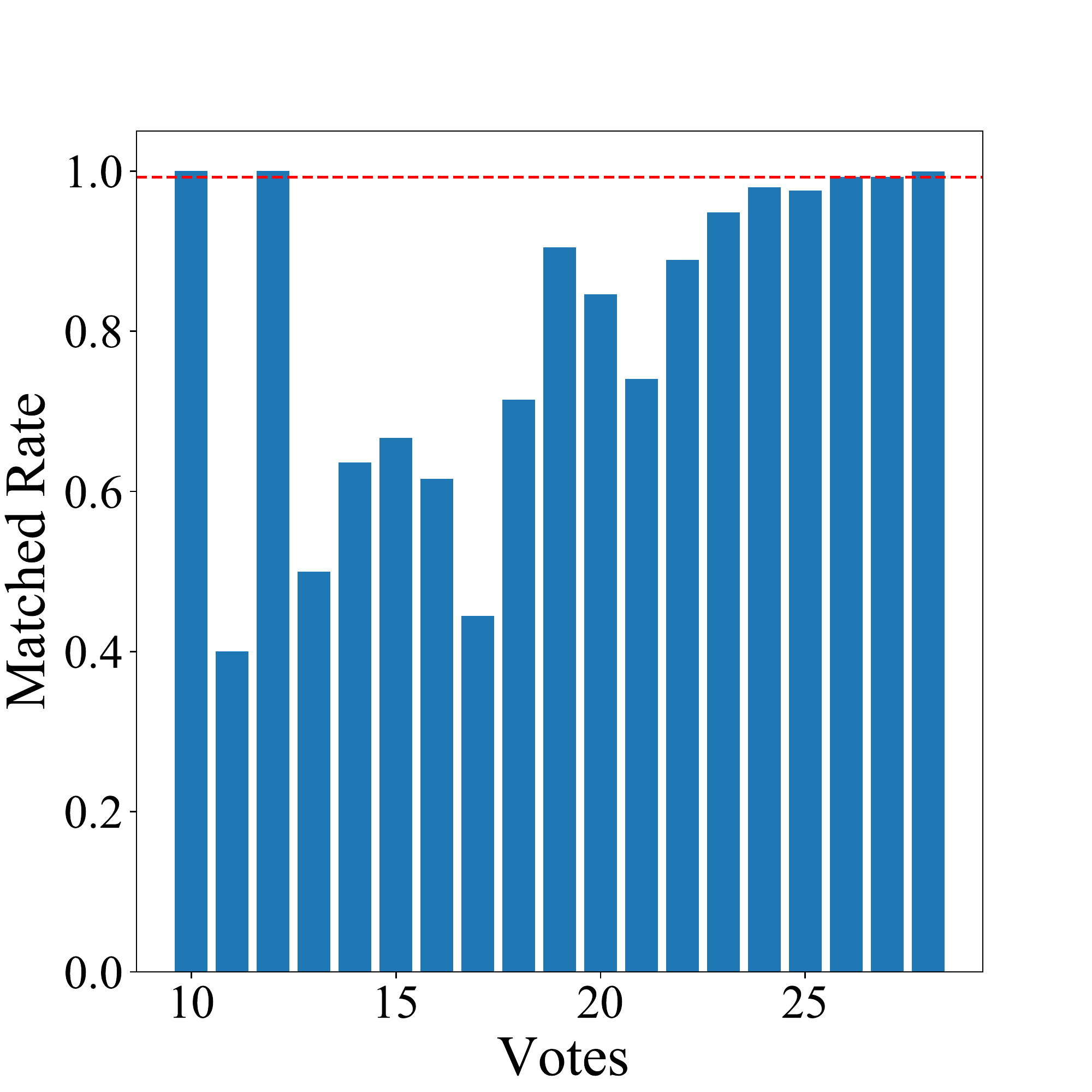}}	
	\subfigure[Fashion-MNIST]{
		\label{DIS3-F} 
		\includegraphics[width=0.15\textwidth]{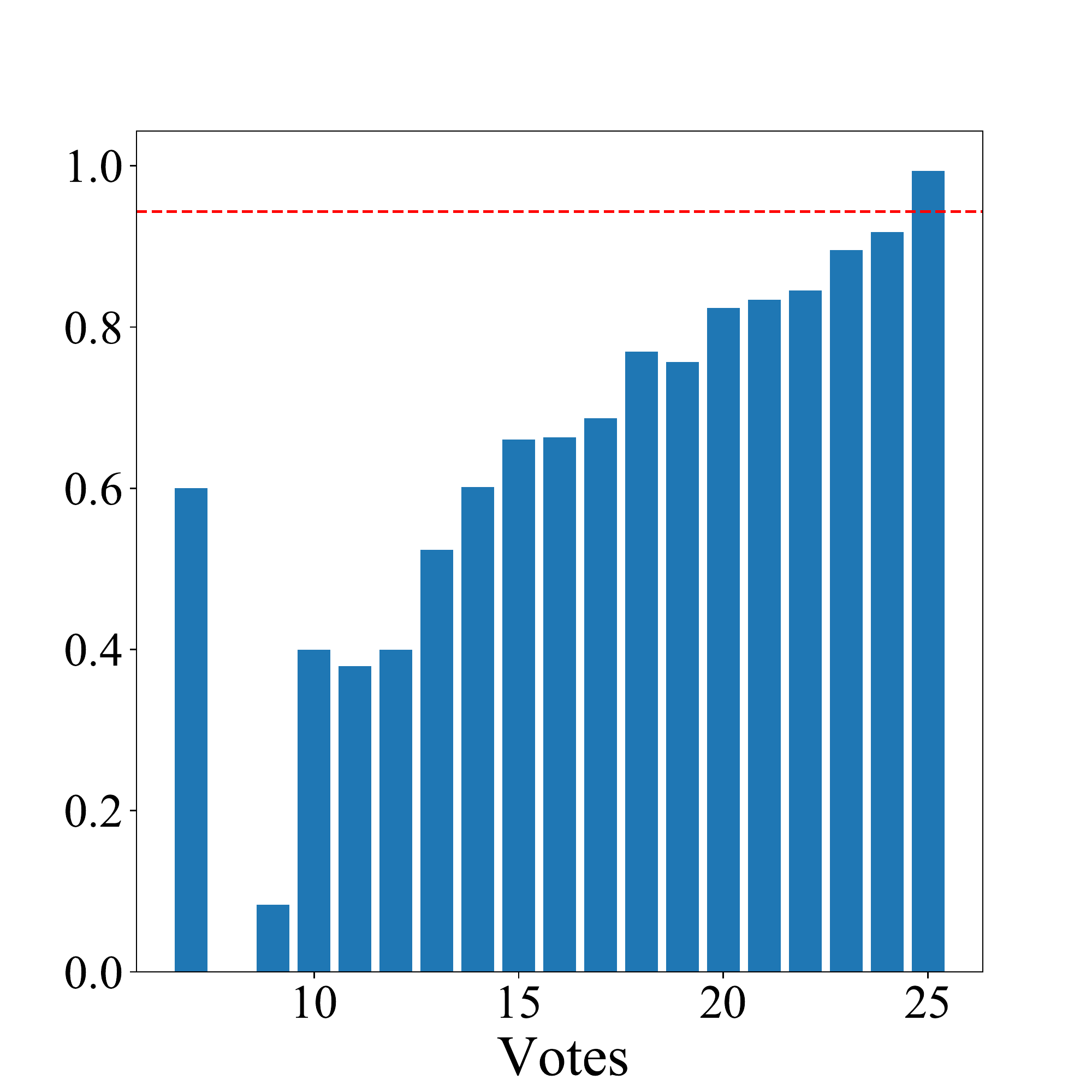}}
	\subfigure[CIFAR-10]{
		\label{DIS3-C} 
		\includegraphics[width=0.15\textwidth]{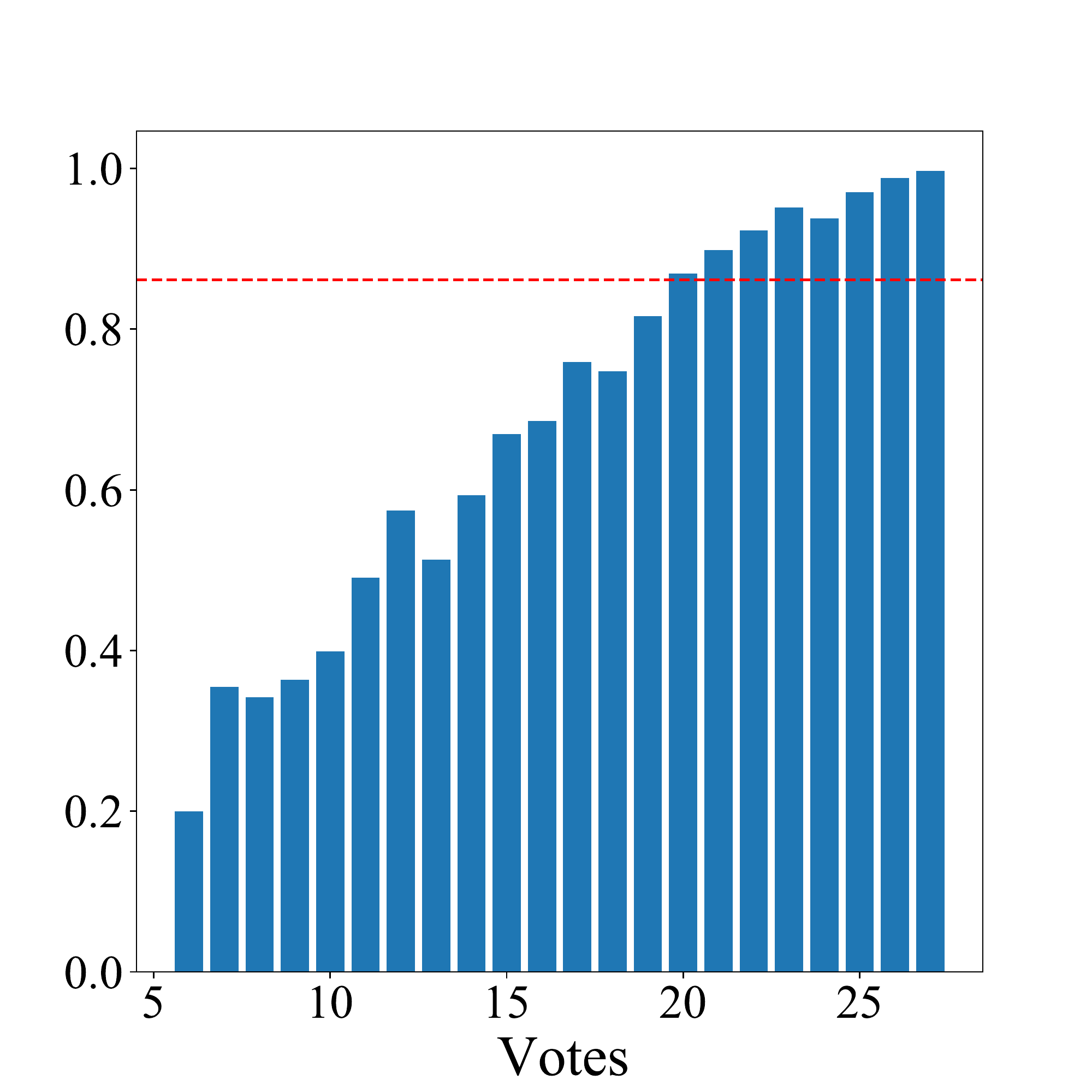}}
	\caption{The histogram of matched rate when the votes changes. The red line represents the matched rate on the entire data set.}
	\label{dis_3_1}
\end{figure}

In this section we will discuss why our algorithm works. In order to illustrate this point, we conduct a two-step analysis. The first is to measure the precision of the majority voting. We compare estimated labels obtained by the majority voting with true labels. Figure~\ref{dis_3_1} shows the matched rate of estimated labels with true labels when the majority voting gets different numbers of votes. It can be seen that as the number of votes obtained increases, the matched rate also rises. In general, the average matched rate of majority voting results with the true labels reaches 0.9924 for MNIST, 0.9433 for Fahion-MNIST, and 0.8613 for CIFAR10, respectively, as shown by the red line in each subfigure. In other words, majority voting is close to the true label, which is the key to explain why our method is effective. This finding leads to an insight for following studies in comparative testing: \textit{the distribution of predicted labels would be helpful to deal with the lack of actual labels, which is a main difficulty in actual testing scenarios due to the limitation of labelling effort}. We encourage following researchers to employ more effective methods to measure the distribution in comparative testing.

\begin{figure}[t]
	\centering
	\subfigure[Spearman]{
		\label{DIS3-S} 
		\includegraphics[width=0.23\textwidth]{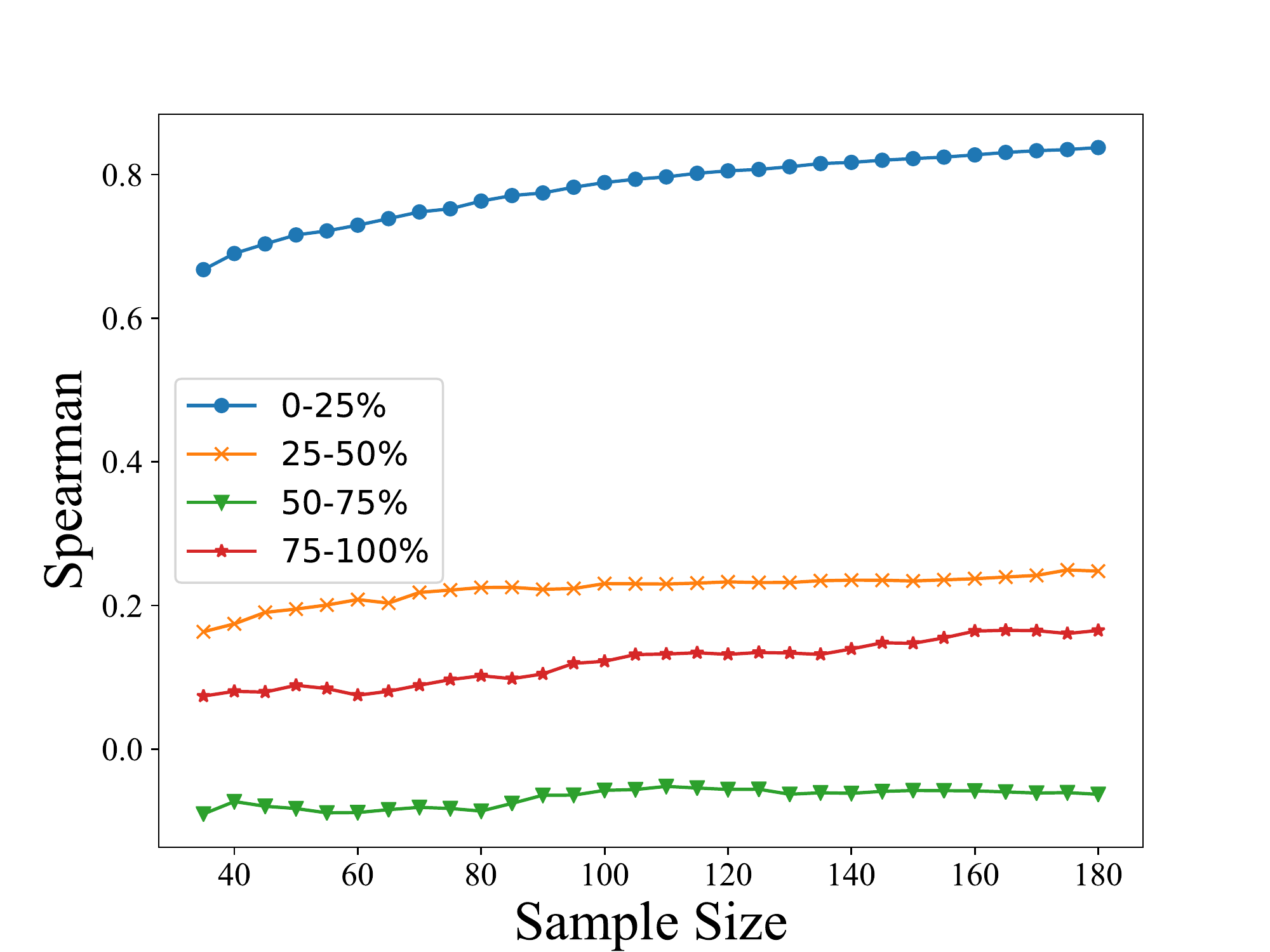}}
	\subfigure[Jaccard]{
		\label{DIS3-J} 
		\includegraphics[width=0.23\textwidth]{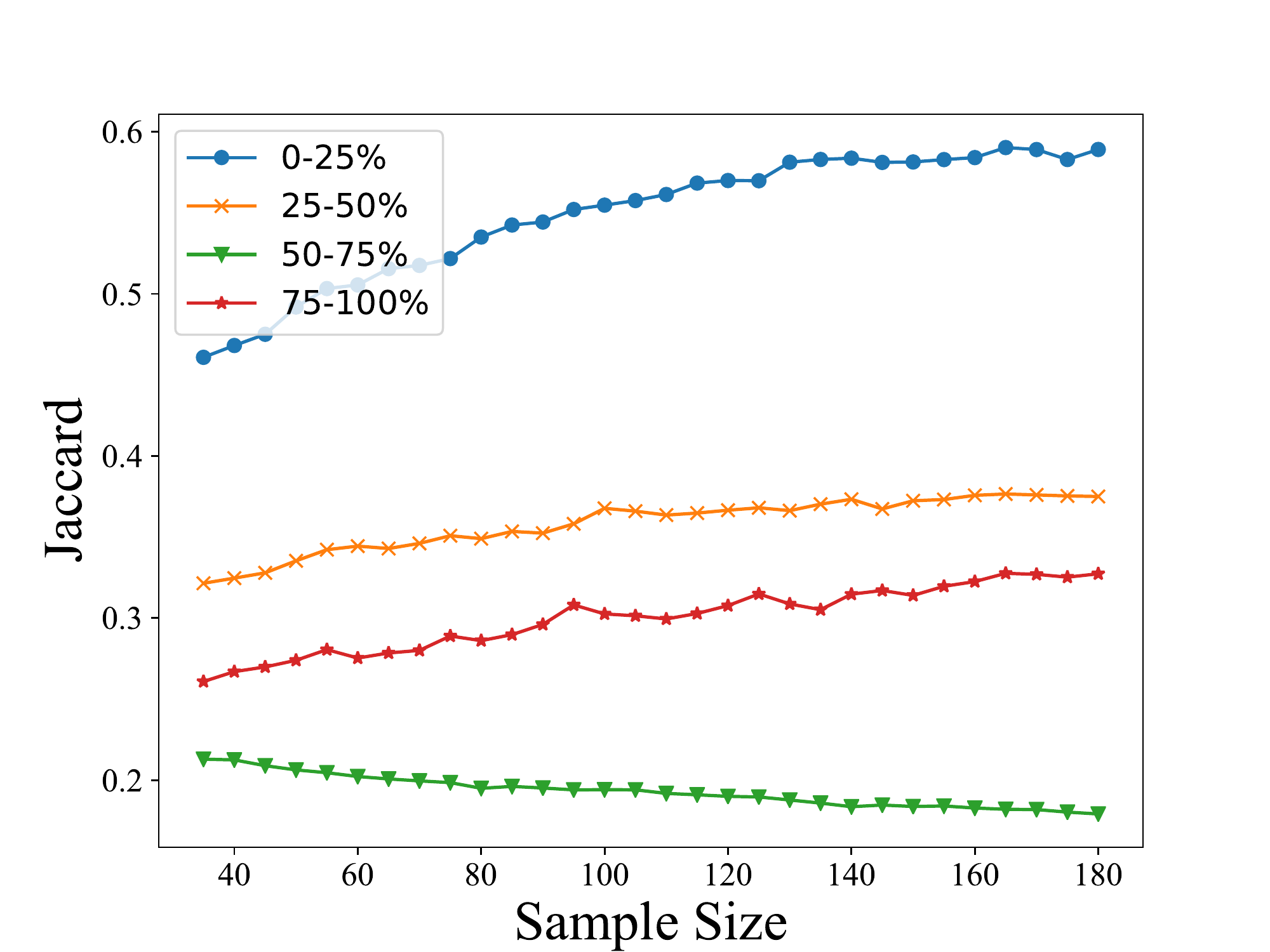}}
	\caption{The graph of four intervals for random sampling (the first 25\%, 25\%-50\%, 50\%-75\%, and 75\%-100\%) to show that the sample discrimination is positively correlated to the ranking performance.}
	\label{dis_3_2}
\end{figure}

In the second step, we analyze whether the sample discrimination is positively correlated to the ranking performance, i.e., whether higher discrimination is more helpful for ranking multiple DL models. We conduct an additional experiment. After sorting the samples according to the discrimination, we randomly select samples in the top 25\%, the 25\%-50\%, the 50\%-75\%, and the 75\%-100\% intervals to observe the results of ranking performance. We take the averages of the three datasets and show them in the Figure~\ref{dis_3_2}, the blue line represents the random sampling in the first 25\% interval, which is the interval used in our experiment. It can be seen that the model ranking effectiveness of random sampling in the first 25\% is significantly better than other intervals. That indicates higher discrimination is more helpful for ranking multiple DL models.

\subsection{The performance when there are fewer models}
\begin{figure}[t]
	\centering
		\label{DIS5} 
		\includegraphics[width=0.28\textwidth]{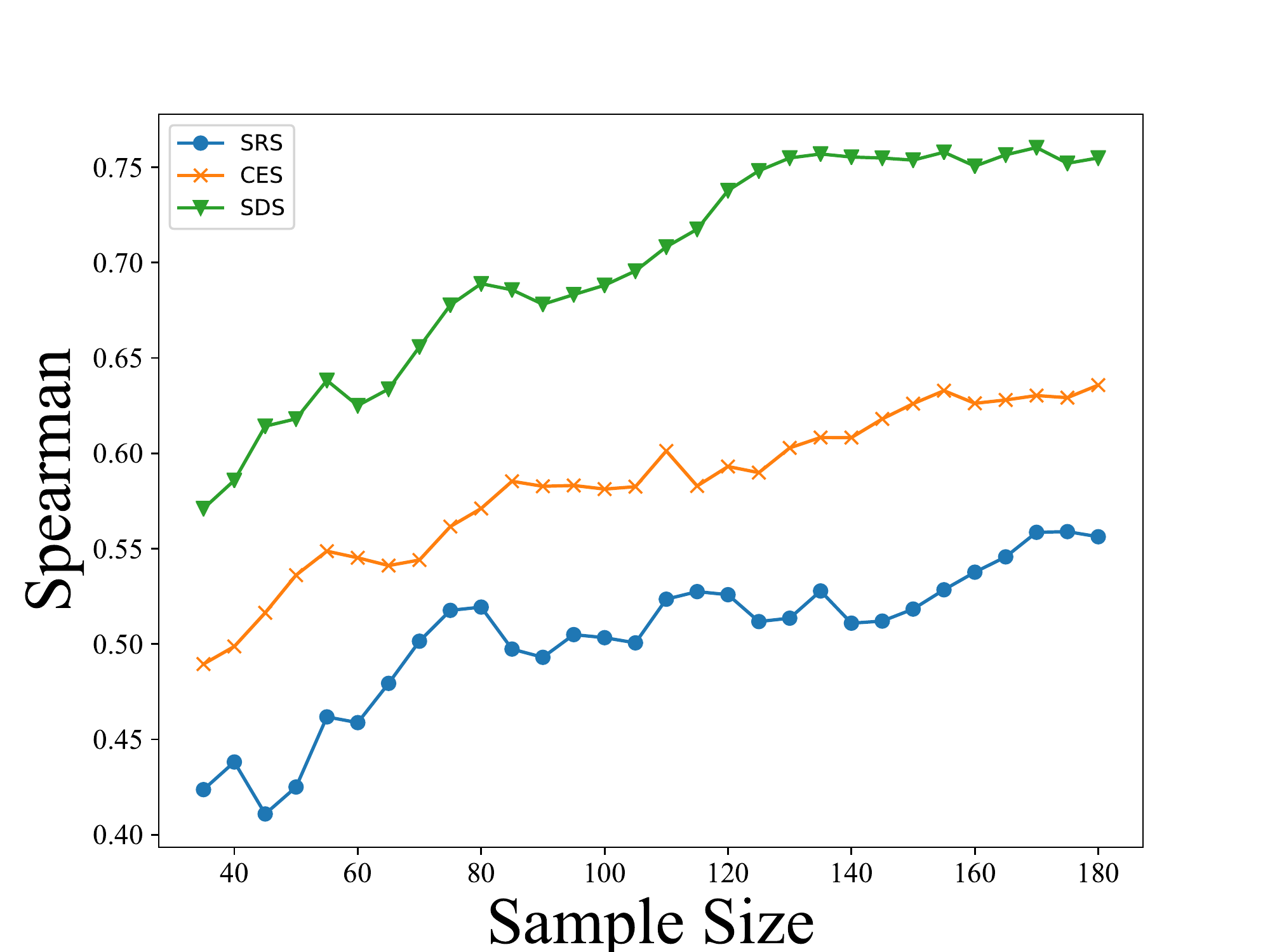}	
	\caption{Comparison result of ranking performance of SDS, SRS, and CES when the number of models is 4.}
	\label{dis_5}
\end{figure}

The previous experiment content is to calculate the ranking performance of the SDS method when the number of models is large (i.e., more than 20 models for a given task). In this section, we report the performance of SDS on the model ranking when there are few models. We have selected four models in each data set to compare the ranking effect of SDS and other baselines. We measure the Spearman coefficient\footnote{As there are only four models, Jaccard coefficient ($k=10$) is not applicable here. We focus on Spearman coefficient.} value when the sample size is from 35 to 180, the experiment was repeated 50 times, and the average results were reported. 
Figure~\ref{dis_5} shows the comparison results of SDS, SRS, and CES, which are the best three methods when the number of models is large. 
Figure~\ref{dis_5} presents the average ranking performance on the three data sets, where the green curve denotes the Spearman coefficient of SDS.

 We observe that SDS can still show superior performance when there are fewer models, which obviously exceeds SRS and CES. To some extent, the above result shows the generalization of the SDS method.

\subsection{The ranking performance when choosing majority voting as true labels}
An intuitive idea is to use the labels obtained by the majority voting as the true labels to measure the accuracy of the models, and then get the ranking performance of the models (see line 21 of Algorithm 1). In this section, we compare this intuitive method with the results of SDS to verify whether the calculation after line 21 in Algorithm 1 really plays a role in model ranking.  

\begin{figure}[t]
	\centering
		\includegraphics[width=0.28\textwidth]{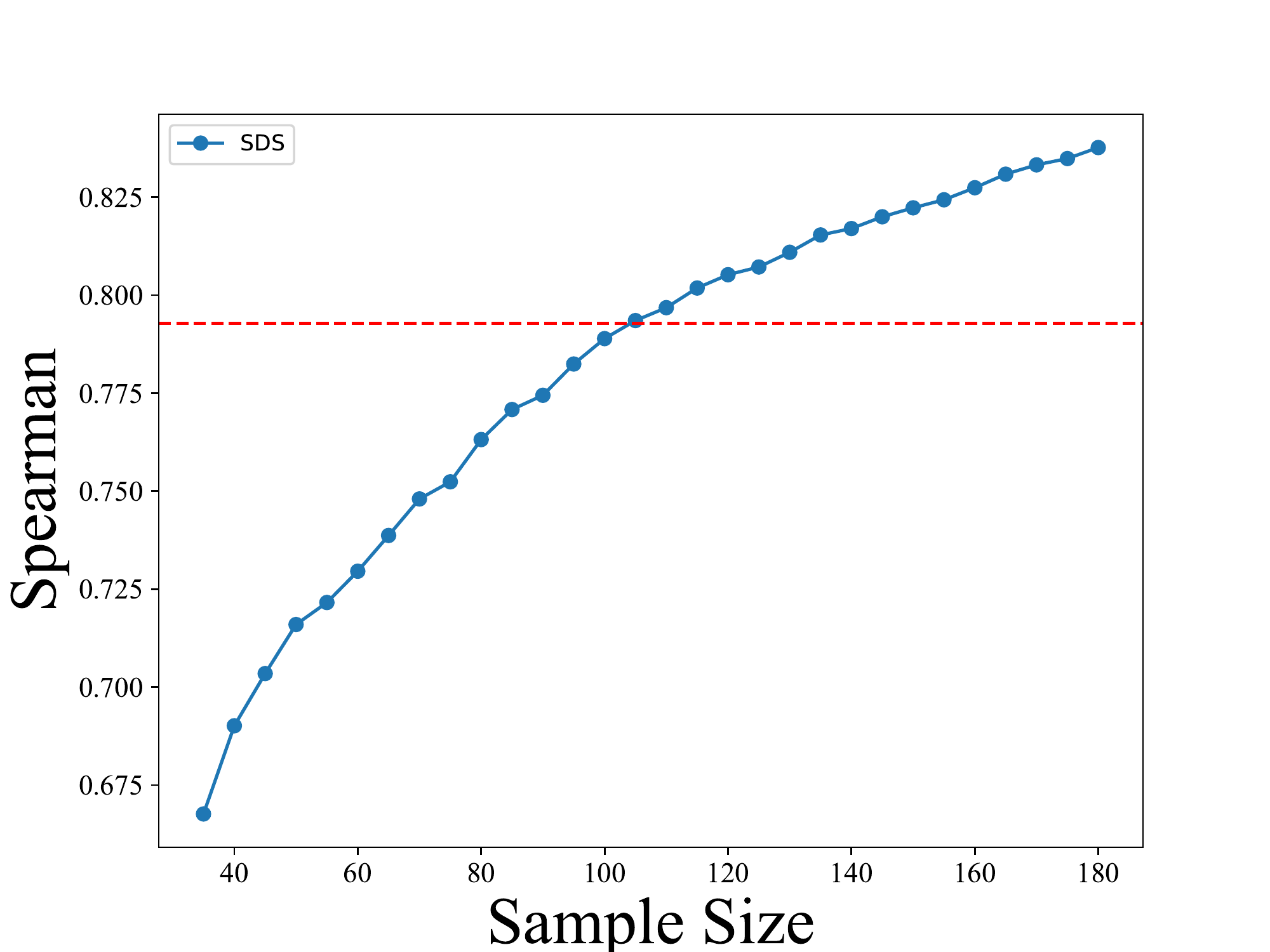}
	\caption{Comparison result of ranking performance of SDS with ranking performance when majority voting is used as the real label.}
	\label{dis_4}
\end{figure}

We show the results of the comparison in Figure~\ref{dis_4}. The blue curve in the figure is the average result of the spearman coefficient on the three data sets that vary with the sample size, and the red line is the ranking result obtained by using the majority voting results as the true labels, which is also the average result of the three data sets. 

It can be seen from Figure~\ref{dis_4} that SDS overcomes majority voting when the sample size is larger than 105, which is roughly about one percent (i.e., 10000*1\%=100) of the total test set. Besides, the curve after this point still shows an upward trend along with the increasing of sampling size. That is to say, the calculation content after line 21 in Algorithm 1 is useful for the model ranking.

\section{THREAD TO VALIDITY}
\label{sec:thread}
The threat to validity is discussed in the following three aspects for our study. 

First, the datasets we select may be a threat. We use three well-known graph classification datasets, which are widely used in many studies, but their complexity is not high. In the future, we will explore on larger and more diverse datasets to validate the effectiveness of our algorithm.

Second, the selection of models in the experiments could become a threat. We try to choose a wide range of models on GitHub, i.e., 28 models for MNIST, 25 for Fashion-MNIST, and 27 for CIFAR-10, respectively, which include multiple DL models with different stars (from a few to tens of thousands) on Github, different model structures, and different accuracies. However, these studied 80 models may not fully cover the real situation. More models are expected in the following studies to validate our results. 

Finally, it may also be a threat to the implementation of the models. As discussed earlier, if the trained model  file is provided in the GitHub repository, we will use it directly, otherwise we will use the provided python code and datasets for training. Due to the difference in the training environment, it may cause the reproduced model to be different from the original one. For new trained models, we compare the accuracies announced in the GitHub repository and actual accuracies, and find that the difference between them is slight.  

\section{RELATEDWORK}
\label{sec:rel}
In this section, we introduce the related work. In the angel of traditional software testing \cite{Frankl1998Evaluating}, on the one side, testing aims to find more bugs, which is called debug testing; on the other side, testing aims to make reliability assessment of software through conditioning, which is called operational testing. 

The main body of current DL testing is to focus on debug testing, i.e., the main aim is to find bugs.  Pei et al. proposed a whitebox framework named DeepXplore, which uses neuron coverage as the standard for DL model testing \cite{DeepXplore2017}. Tian et al. implemented a tool named DeepTest to simulate the real world to help find behaviors that may cause accidents for DNN-driven vehicles \cite{tian2018deeptest}. Zhang et al. proposed unsupervised framework for DNN named DeepRoad, and utilized GANs and metamorphic testing to test the inconsistent behaviors in self-driving car \cite{DeepRoad2018}. Xie et al. proposed a coverage-based framework named DeepHunter which used metamorphic mutation to help find defects for DNNs \cite{DBLP:conf/issta/XieMJXCLZLYS19}. Sun et al. presented an approach named TransRepair to help machine translation systems test and repair inconsistency bugs \cite{DBLP:journals/corr/abs-1910-02688}. Ma et al. proposed a set of testing criteria named DeepGauge for measuring the testing adequacy of DNNs \cite{DeepGauge2018}. Ma et al. proposed DeepCT, which applied the idea of combinatorial testing to DL testing, and produced a series of combinatorial testing criteria for DL systems \cite{DBLP:conf/wcre/MaJXLLLZ19}. Tian et al. developed a technique called DeepInspect, which can detect the confusion and bias errors based class for image classification \cite{DBLP:journals/corr/abs-1905-07831}. Lee et al. presented a white-box testing approach named ADAPT, which used an adaptive neuron selection strategy to find adversarial inputs \cite{DBLP:conf/issta/LeeCLO20}.

Meanwhile, researchers have focused on the other aspects of DL testing. Li et al. proposed an effective operational testing technique to estimate the accuracy of the DL model by constructing probabilistic models for the distribution of testing contexts \cite{Li2019}. To evaluate the quality of test data, Ma et al. applied the mutation framework to DL systems, and proposed a technique named DeepMutation \cite{DeepMutation}. Zhou et al. proposed a testing approach faced the systematic physical world called DeepBillboard, which is aimed to generate adversarial test more robust \cite{DBLP:journals/corr/abs-1812-10812}. Gerasimou et al. proposed a systematic testing approach named DeepImportance, which is mixed with an Importance-Driven (IDC) test adequacy criterion to support more robust DL systems \cite{DBLP:journals/corr/abs-2002-03433}.

\section{CONCLUSION}
\label{sec:con}

The boom of DL technology leads to the reuse of DL models, which expedites the emergence of a new testing scenario \textit{comparative testing}, where testers may encounter multiple DL models with the same functionality as candidates to accomplish a specific task, and testers are expected to rank them to choose the more suitable models in the testing contexts. Due to the limitation of labeling effort, this testing scenario brings out a new problem of DL testing: \textit{ranking multiple DL models under limited labeling efforts}. 

To tackle this problem, we propose a novel algorithm named \textbf{S}ample \textbf{D}iscrimination based \textbf{S}election (\textbf{SDS}) to measure the sample discrimination and select samples with higher discrimination. We evaluate our approach on three widely-used image datasets and 80 DL models. Our results lead us to conclude that SDS is an effective and efficient sample selection method for comparative testing to rank multiple DL models.

Finally, we would like to emphasize that we do not seek to
claim the advantage of our method SDS. Instead, the key messages are that (a) a new testing scenario \textit{comparative testing} is introduced by our paper, where the testing aims are much different with the current DL testing, i.e., debug/operational testing; (b) the new testing scenario brings out the new testing challenge \textit{ranking multiple DL models under limited labeling efforts}; (c) our proposed method SDS leads to the insight which would be helpful for the following researchers.

\section{REPEATABILITY}
\label{rep}

We provide all datasets and code used to conduct
this study at \url{https://github.com/Testing-Multiple-DL-Models/SDS}.

\section*{Acknowledgements} 
The work is supported by National Key R\&D Program of China (Grant No. 2018YFB1003901) and the National Natural Science Foundation of China (Grant No. 61872177, 61832009, 61772259, 61772263, and 61932012). We thank the anonymous referees for their helpful comments on this paper. 

\balance
\bibliographystyle{abbrv}
\bibliography{ref1,ref2} 

\begin{thebibliography}{10}

\bibitem{beller2015and}
M.~Beller, G.~Gousios, A.~Panichella, and A.~Zaidman.
\newblock When, how, and why developers (do not) test in their ides.
\newblock In {\em Proceedings of the 2015 10th Joint Meeting on Foundations of
  Software Engineering}, pages 179--190, 2015.

\bibitem{rank_sum}
L.~D. Capitani and D.~D. Martini.
\newblock On stochastic orderings of the wilcoxon rank sum test statistic with
  applications to reproducibility probability estimation testing.
\newblock {\em Statistics and Probability Letters}, 81(8):937--946, 2011.

\bibitem{ebel1954procedures}
R.~L. Ebel.
\newblock Procedures for the analysis of classroom tests.
\newblock {\em Educational and Psychological Measurement}, 14(2):352--364,
  1954.

\bibitem{farabet2012scene}
C.~Farabet, C.~Couprie, L.~Najman, and Y.~LeCun.
\newblock Scene parsing with multiscale feature learning, purity trees, and
  optimal covers.
\newblock {\em arXiv preprint arXiv:1202.2160}, 2012.

\bibitem{feng2020}
Y.~Feng, Q.~Shi, X.~Gao, J.~Wan, C.~Fang, and Z.~Chen.
\newblock Deepgini: Prioritizing massive tests to enhance the robustness of
  deep neural networks.
\newblock In {\em Proceedings of the 29th ACM SIGSOFT International Symposium
  on Software Testing and Analysis}, ISSTA 2020, page 177–188, New York, NY,
  USA, 2020. Association for Computing Machinery.

\bibitem{Frankl1998Evaluating}
Frankl, Phyllis, G., Hamlet, Richard, G., Littlewood, Bev, Strigini, and
  Lorenzo.
\newblock Evaluating testing methods by delivered reliability.
\newblock {\em IEEE Transactions on Software Engineering}, 1998.

\bibitem{geiger2012we}
A.~Geiger, P.~Lenz, and R.~Urtasun.
\newblock Are we ready for autonomous driving? the kitti vision benchmark
  suite.
\newblock In {\em 2012 IEEE Conference on Computer Vision and Pattern
  Recognition}, pages 3354--3361. IEEE, 2012.

\bibitem{DBLP:journals/corr/abs-2002-03433}
S.~Gerasimou, H.~F. Eniser, A.~Sen, and A.~Cakan.
\newblock Importance-driven deep learning system testing.
\newblock {\em CoRR}, abs/2002.03433, 2020.

\bibitem{hinton2012deep}
G.~Hinton, L.~Deng, D.~Yu, G.~E. Dahl, A.-r. Mohamed, N.~Jaitly, A.~Senior,
  V.~Vanhoucke, P.~Nguyen, T.~N. Sainath, et~al.
\newblock Deep neural networks for acoustic modeling in speech recognition: The
  shared views of four research groups.
\newblock {\em IEEE Signal processing magazine}, 29(6):82--97, 2012.

\bibitem{10.1145/3243734.3243757}
Y.~Ji, X.~Zhang, S.~Ji, X.~Luo, and T.~Wang.
\newblock Model-reuse attacks on deep learning systems.
\newblock In {\em Proceedings of the 2018 ACM SIGSAC Conference on Computer and
  Communications Security}, CCS '18, page 349–363, New York, NY, USA, 2018.
  Association for Computing Machinery.

\bibitem{Surprise_Adequacy}
J.~Kim, R.~Feldt, and S.~Yoo.
\newblock Guiding deep learning system testing using surprise adequacy.
\newblock In {\em Proceedings of the 41st International Conference on Software
  Engineering}, ICSE '19, pages 1039--1049, Piscataway, NJ, USA, 2019. IEEE
  Press.

\bibitem{Krizhevsky2017ImageNet}
Krizhevsky, Alex, Sutskever, Ilya, Hinton, and E.~Geoffrey.
\newblock Imagenet classification with deep convolutional neural networks.
\newblock {\em Communications of the ACM}, 2017.

\bibitem{lecun2015deep}
Y.~LeCun, Y.~Bengio, and G.~Hinton.
\newblock Deep learning.
\newblock {\em nature}, 521(7553):436, 2015.

\bibitem{MNIST}
Y.~LeCun and C.~Cortes.
\newblock The mnist database of handwritten digits.
\newblock \url{http://yann.lecun.com/exdb/mnist/}, 2019.
\newblock Accessed May 4, 2019.

\bibitem{DBLP:conf/issta/LeeCLO20}
S.~Lee, S.~Cha, D.~Lee, and H.~Oh.
\newblock Effective white-box testing of deep neural networks with adaptive
  neuron-selection strategy.
\newblock In S.~Khurshid and C.~S. Pasareanu, editors, {\em {ISSTA} '20: 29th
  {ACM} {SIGSOFT} International Symposium on Software Testing and Analysis,
  Virtual Event, USA, July 18-22, 2020}, pages 165--176. {ACM}, 2020.

\bibitem{Li2019}
Z.~Li, X.~Ma, C.~Xu, C.~Cao, J.~Xu, and J.~L\"{u}.
\newblock Boosting operational dnn testing efficiency through conditioning.
\newblock In {\em Proceedings of the 2019 27th ACM Joint Meeting on European
  Software Engineering Conference and Symposium on the Foundations of Software
  Engineering}, ESEC/FSE 2019, page 499–509, New York, NY, USA, 2019.
  Association for Computing Machinery.

\bibitem{liu2018connecting}
Y.~Liu, Y.~Li, J.~Guo, Y.~Zhou, and B.~Xu.
\newblock Connecting software metrics across versions to predict defects.
\newblock In {\em 2018 IEEE 25th International Conference on Software Analysis,
  Evolution and Reengineering (SANER)}, pages 232--243. IEEE, 2018.

\bibitem{DBLP:conf/wcre/MaJXLLLZ19}
L.~Ma, F.~Juefei{-}Xu, M.~Xue, B.~Li, L.~Li, Y.~Liu, and J.~Zhao.
\newblock Deepct: Tomographic combinatorial testing for deep learning systems.
\newblock In X.~Wang, D.~Lo, and E.~Shihab, editors, {\em 26th {IEEE}
  International Conference on Software Analysis, Evolution and Reengineering,
  {SANER} 2019, Hangzhou, China, February 24-27, 2019}, pages 614--618. {IEEE},
  2019.

\bibitem{DeepGauge2018}
L.~Ma, F.~Juefei{-}Xu, F.~Zhang, J.~Sun, M.~Xue, B.~Li, C.~Chen, T.~Su, L.~Li,
  Y.~Liu, J.~Zhao, and Y.~Wang.
\newblock Deepgauge: multi-granularity testing criteria for deep learning
  systems.
\newblock In {\em Proceedings of the 33rd {ACM/IEEE} International Conference
  on Automated Software Engineering, {ASE} 2018, Montpellier, France, September
  3-7, 2018}, pages 120--131, 2018.

\bibitem{DeepMutation}
L.~{Ma}, F.~{Zhang}, J.~{Sun}, M.~{Xue}, B.~{Li}, F.~{Juefei-Xu}, C.~{Xie},
  L.~{Li}, Y.~{Liu}, J.~{Zhao}, and Y.~{Wang}.
\newblock Deepmutation: Mutation testing of deep learning systems.
\newblock In {\em 2018 IEEE 29th International Symposium on Software
  Reliability Engineering (ISSRE)}, pages 100--111, Oct 2018.

\bibitem{WTL}
J.~{Nam}, W.~{Fu}, S.~{Kim}, T.~{Menzies}, and L.~{Tan}.
\newblock Heterogeneous defect prediction.
\newblock {\em IEEE Transactions on Software Engineering}, 44(9):874--896, Sep.
  2018.

\bibitem{cifar_10}
N.Krizhevsky, H.Vinod, C.Geoffrey, M.Papadakis, and A.Ventresque.
\newblock The cifar-10 dataset.
\newblock \url{http://www.cs.toronto.edu/~kriz/cifar.html}.
\newblock Accessed May 4, 2019.

\bibitem{DeepXplore2017}
K.~Pei, Y.~Cao, J.~Yang, and S.~Jana.
\newblock Deepxplore: Automated whitebox testing of deep learning systems.
\newblock In {\em Proceedings of the 26th Symposium on Operating Systems
  Principles, Shanghai, China, October 28-31, 2017}, pages 1--18, 2017.

\bibitem{pouyanfar2018survey}
S.~Pouyanfar, S.~Sadiq, Y.~Yan, H.~Tian, Y.~Tao, M.~P. Reyes, M.-L. Shyu, S.-C.
  Chen, and S.~Iyengar.
\newblock A survey on deep learning: Algorithms, techniques, and applications.
\newblock {\em ACM Computing Surveys (CSUR)}, 51(5):1--36, 2018.

\bibitem{cliff_delta}
J.~{Romano}, J.~D. {Kromrey}, J.~{Coraggio}, J.~{Skowronek}, and L.~{Devine}.
\newblock Exploring methods for evaluating group differences on the nsse and
  other surveys: Are the t-test and cohen's d indices the most appropriate
  choices.
\newblock In {\em In annual meeting of the Southern Association for
  Institutional Research}, 2006.

\bibitem{sagi2018ensemble}
O.~Sagi and L.~Rokach.
\newblock Ensemble learning: A survey.
\newblock {\em Wiley Interdisciplinary Reviews: Data Mining and Knowledge
  Discovery}, 8(4):e1249, 2018.

\bibitem{sculley2015hidden}
D.~Sculley, G.~Holt, D.~Golovin, E.~Davydov, T.~Phillips, D.~Ebner,
  V.~Chaudhary, M.~Young, J.-F. Crespo, and D.~Dennison.
\newblock Hidden technical debt in machine learning systems.
\newblock In {\em Advances in neural information processing systems}, pages
  2503--2511, 2015.

\bibitem{9286133}
W.~{Shen}, Y.~{Li}, L.~{Chen}, Y.~{Han}, Y.~{Zhou}, and B.~{Xu}.
\newblock Multiple-boundary clustering and prioritization to promote neural
  network retraining.
\newblock In {\em 2020 35th IEEE/ACM International Conference on Automated
  Software Engineering (ASE)}, pages 410--422, 2020.

\bibitem{silver2016mastering}
D.~Silver, A.~Huang, C.~J. Maddison, A.~Guez, L.~Sifre, G.~Van Den~Driessche,
  J.~Schrittwieser, I.~Antonoglou, V.~Panneershelvam, M.~Lanctot, et~al.
\newblock Mastering the game of go with deep neural networks and tree search.
\newblock {\em nature}, 529(7587):484--489, 2016.

\bibitem{DBLP:journals/corr/abs-1910-02688}
Z.~Sun, J.~M. Zhang, M.~Harman, M.~Papadakis, and L.~Zhang.
\newblock Automatic testing and improvement of machine translation.
\newblock {\em CoRR}, abs/1910.02688, 2019.

\bibitem{thongtanunam2020review}
P.~Thongtanunam and A.~E. Hassan.
\newblock Review dynamics and their impact on software quality.
\newblock {\em IEEE Transactions on Software Engineering}, 2020.

\bibitem{tian2018deeptest}
Y.~Tian, K.~Pei, S.~Jana, and B.~Ray.
\newblock Deeptest: Automated testing of deep-neural-network-driven autonomous
  cars.
\newblock In {\em Proceedings of the 40th international conference on software
  engineering}, pages 303--314, 2018.

\bibitem{DBLP:journals/corr/abs-1905-07831}
Y.~Tian, Z.~Zhong, V.~Ordonez, and B.~Ray.
\newblock Testing deep neural network based image classifiers.
\newblock {\em CoRR}, abs/1905.07831, 2019.

\bibitem{xiao2017fashion}
H.~Xiao, K.~Rasul, and R.~Vollgraf.
\newblock Fashion-mnist: a novel image dataset for benchmarking machine
  learning algorithms.
\newblock {\em arXiv preprint arXiv:1708.07747}, 2017.

\bibitem{DBLP:conf/issta/XieMJXCLZLYS19}
X.~Xie, L.~Ma, F.~Juefei{-}Xu, M.~Xue, H.~Chen, Y.~Liu, J.~Zhao, B.~Li, J.~Yin,
  and S.~See.
\newblock Deephunter: a coverage-guided fuzz testing framework for deep neural
  networks.
\newblock In D.~Zhang and A.~M{\o}ller, editors, {\em Proceedings of the 28th
  {ACM} {SIGSOFT} International Symposium on Software Testing and Analysis,
  {ISSTA} 2019, Beijing, China, July 15-19, 2019}, pages 146--157. {ACM}, 2019.

\bibitem{You:2017}
Y.~You, A.~Bulu{\c{c}}, and J.~Demmel.
\newblock Scaling deep learning on gpu and knights landing clusters.
\newblock In {\em Proceedings of the International Conference for High
  Performance Computing, Networking, Storage and Analysis}, SC '17, pages
  9:1--9:12, New York, NY, USA, 2017. ACM.

\bibitem{DeepRoad2018}
M.~Zhang, Y.~Zhang, L.~Zhang, C.~Liu, and S.~Khurshid.
\newblock Deeproad: Gan-based metamorphic testing and input validation
  framework for autonomous driving systems.
\newblock In {\em Proceedings of the 33rd {ACM/IEEE} International Conference
  on Automated Software Engineering, {ASE} 2018, Montpellier, France, September
  3-7, 2018}, pages 132--142, 2018.

\bibitem{DBLP:journals/corr/abs-1812-10812}
H.~Zhou, W.~Li, Y.~Zhu, Y.~Zhang, B.~Yu, L.~Zhang, and C.~Liu.
\newblock Deepbillboard: Systematic physical-world testing of autonomous
  driving systems.
\newblock {\em CoRR}, abs/1812.10812, 2018.

\bibitem{zhou2015metamorphic}
Z.~Q. Zhou, S.~Xiang, and T.~Y. Chen.
\newblock Metamorphic testing for software quality assessment: A study of
  search engines.
\newblock {\em IEEE Transactions on Software Engineering}, 42(3):264--284,
  2015.

\end{thebibliography}

\end{document}